\documentclass[aps,prd,showpacs,twocolumn]{revtex4}

\usepackage{amsmath,amssymb,graphicx,afterpage}

\newcommand{\be}{\begin{equation}}
\newcommand{\ee}{\end{equation}}
\newcommand{\bea}{\begin{eqnarray}}
\newcommand{\eea}{\end{eqnarray}}
\newcommand{\pr}{\partial}
\newcommand{\cI}{\mathcal{C}_I}
\newcommand{\non}{\nonumber}

\begin{document}
\title{How Current Loops and Solenoids Curve Space-time}

\author{Andr\'e F\"uzfa}
\email{andre.fuzfa@unamur.be}
\affiliation{Namur Center for Complex systems (naXys),\\ University of Namur, Belgium}

\date{\today}

\begin{abstract}
\noindent 
The curved space-time around current loops and solenoids carrying arbitrarily large steady electric currents is obtained from the numerical resolution of the coupled Einstein-Maxwell equations in cylindrical symmetry. The artificial gravitational field associated to the generation of a magnetic field  produces gravitational redshift of photons and deviation of light. Null geodesics in the curved space-time of current loops and solenoids are also presented. We finally propose an experimental setup, achievable with current technology of superconducting coils, that produces a phase shift of light of the same order of magnitude than astrophysical signals in ground-based gravitational wave observatories.
 \end{abstract}

\maketitle

\section{Introduction}
\noindent 
Somehow, studying gravity is a contemplative activity: physicists restrict themselves to the study of \textit{natural}, pre-existing, sources of gravitation. Generating \textit{artificial gravitational fields}, that could be switched on or off at will, is a question captured or left to science-fiction. 

However, the equivalence principle, at the very heart of Einstein's general relativity, states that all types of energy produce and undergo gravitation in the same way. The most widespread source of gravitation is the inertial mass, which produces \textit{permanent} gravitational fields. At the opposite, electromagnetic fields could be used to generate \textit{artificial}, or human-made, gravitational fields, that could be switched on or off at will, depending whether their electromagnetic progenitors are
present or not. \\
\\
The equivalence principle actually implies that one also generates gravitational fields when generating electromagnetic fields. 
%% response to referee comment (a)
However, since the gravitational strength is extremely small compared to the one of the electromagnetic force\footnote{Their strength differs by a factor of about $10^{-40}$ in a hydrogen atom.}, large electromagnetic fields will only produce tiny space-time deformations.
Yet, electromagnetic fields
do curve space-time. 
Therefore, general relativity predicts that light and more generally electrically neutral massive particles are deflected by electromagnetic fields, although they do not feel the classical Lorentz force. This effect does not require new exotic physics and it might serve in the future to build new tests of the equivalence principle in the laboratory. In experimental gravity, the permanent gravitational fields involved cannot be withdrawn completely. At the opposite, the gravitational fields generated by electromagnetic fields can be switched off: their experimental search can therefore be done by comparing measurements made in presence and in absence of electromagnetic fields. 
%% response to referee comment (a)
However, due to the weakness of the gravitational interaction, even the strongest magnetic fields humans can currently generate will only produce tiny space-time deformations. Detecting them would constitute a true experimental challenge which we glimpse at in this paper. Such a detection would nevertheless open the way to new laboratory tests of  the equivalence principle.\\
\\
The so-called Einstein-Maxwell (EM) equations regroup the classical field theories of general relativity and electromagnetism in a covariant way, although without truly unifying them. 
In some sense, the idea of gravitational field generation from a magnetic field can be attributed to Levi-Civita whose early analytical work \cite{levicivita} describes the curvature of space-time completely filled by a uniform magnetic field. Subsequent works have established analytical solutions for space-time around an infinitely long straight wire carrying steady current \cite{mukherji,witten,bonnor54}. The main problem of these analytical solutions is that the associated metric is not asymptotically flat, due to the infinitely large current distribution, which makes these analytical solutions of poor interest for practical applications. Overcoming this problem requires considering current distributions of finite extent such as current loops and solenoids. Asymptotic space-time around current loops carrying steady current has been studied in \cite{bonnor54}. An attempt to derive the full solution of EM equations around the current loop was realized a bit later \cite{bonnor60}, however this attempt lead to an unphysical solution due to an oversimplifying assumption. The case of an infinitely long solenoid was considered in \cite{ivanov} but only for weak perturbations of the metric in linearized general relativity. Therefore, the solutions of the full non-linear EM equations sourced by the steady currents carried by loops and finite solenoids remained so far unexplored until now. \\
\\
In this paper, we present two important results: (1) how space-time is curved around current loops and solenoids carrying \textit{arbitrarily large} electric currents and (2) how the consequent deviation of light could be detected. The structure of this paper is as follows. In section II and III, we develop the numerical resolution of coupled  EM equations in cylindrical symmetry. We then present the trajectories of light in curved space-times around current loops and solenoids in section IV. In section V, we finally propose an experimental set-up, based on a modification of interferometers used for the search for gravitational waves and current technology of superconducting electro-magnets, that produces a detectable artificial space-time curvature.
We conclude by emphasizing the importance of the effect presented here: the deflexion of light in the curved space-time of an electromagnet opens the way to new experimental tests of Einstein's equivalence principle.

\section{Einstein-Maxwell equations for the current loop and the solenoid}
\subsection{Field equations}
The EM system models the interaction of gravitation and electromagnetism 
by their juxtaposition in the following coupled tensorial field equations (in S.I. units\footnote{The relevant fundamental constants of the EM system are $G$ as Newton's constant, $c$ as the speed of light and $\mu_0$ as the (vacuum) magnetic permeability.}):
\bea
R_{\mu\nu}&=&-\frac{8\pi G}{c^4} T^{\left(\rm em\right)}_{\mu\nu} \; \label{einstein}\\
\nabla_\mu F^{\mu\nu}&=&\mu_0 J^\nu\; \label{maxwell}
\eea
where $T^{\left(\rm em\right)}_{\mu\nu}=-\frac{1}{\mu_0}\left(g^{\alpha\beta}F_{\mu\alpha}F_{\nu\beta}-\frac{1}{4}g_{\mu\nu}F_{\alpha\beta}F^{\alpha\beta}\right)$ is the Maxwell stress-energy tensor, $F_{\mu\nu}=\partial_\mu A_\nu-\partial_\nu A_\mu$
is the Faraday tensor of the electromagnetic field, $g_{\mu\nu}$ and $A_\mu$ are the metric and the four-vector potential, i.e. the fundamental fields describing gravitation and electromagnetism respectively, $R_{\mu\nu}$ is the Ricci tensor and $J^\nu$ the four-current density. 

Space-time is therefore curved by the energy of the electromagnetic field as ruled by Einstein equations of general relativity (\ref{einstein}). In the same time, the electromagnetic field propagates in the non-trivial background it generated through Einstein's equations (\ref{einstein}), and this propagation is described by the covariant Maxwell equations in curved space-time (\ref{maxwell}). Because we are interested in the $additionnal$ gravitational field that is produced by the magnetic field, we neglect the mass of the current carriers and the electric wires\footnote{Experimentally, any effect of the mass carriers can easily be handled by calibrating the experiment in the absence of electric currents.}. 

Since current loops and solenoids possess one axis of symmetry, we choose the so-called Weyl gauge \cite{bonnor54,bonnor60,weyl} for the metric field:
\be
ds^2=c^2 e^{\rho(r,z)}dt^2-e^{\lambda(r,z)}\left(dr^2+dz^2\right)-e^{-\rho(r,z)}r^2d\varphi^2\cdot \label{weyl}
\ee
In this symmetry, the vector potential $A_\mu$ trivially reduces to one non-vanishing magnetic component $A_\varphi=a(r,z)/r$.
The advantage of the Weyl gauge (\ref{weyl}) is that the equations of motion directly exhibit usual Laplacian operators on flat background with cylindrical coordinates. Indeed, Eqs.
(\ref{einstein}) and (\ref{maxwell}) using Eq.(\ref{weyl}) now read (see also \cite{bonnor60}):
\bea
\nabla_{(r,z)}^2\rho&=&\frac{8\pi G}{c^4\mu_0}\frac{e^\rho}{r^2}\left(\left(\pr_r a\right)^2+\left(\pr_z a\right)^2\right)\label{rho1}\\
\nabla_{(r,z)}^2\lambda+\left(\pr_z\rho\right)^2 &=& \frac{8\pi G}{c^4\mu_0}\frac{e^\rho}{r^2}\left(\left(\pr_r a\right)^2-\left(\pr_z a\right)^2\right)\label{lam1}\\
\pr_z\lambda +\pr_z\rho&=&r\pr_r\rho\pr_z\rho+\frac{16\pi G}{c^4\mu_0}\frac{e^\rho}{r}\pr_r a\pr_z a\label{lam1bis}\\
\nabla_{(r,z)}^2 a-\frac{2}{r}\pr_r a&=&-\left(\pr_r a\pr_r \rho+\pr_z a\pr_z \rho\right)-r\mu_0 J \label{maxwell2}
\eea
where $\nabla^2_{(r,z)}=\pr_r^2 +\frac{1}{r}\pr_r+\pr_z^2$ the usual Laplacian on flat space in cylindrical coordinates and where $J$
is the angular component of the current density. In \cite{bonnor60}, the term $\left(\pr_r a\pr_r \rho+\pr_z a\pr_z \rho\right)$ in Eq.(\ref{maxwell2}) was arbitrarily set to zero to provide an analytical solution. We do not restrict ourselves here to this arbitrary reducing condition and provide the full solution numerically.

Eq.(\ref{maxwell2}) is the Maxwell equation on a curved space-time described by cylindrical coordinates. For a flat Minkowski background $\rho=\lambda=0$, we have that  the non-relativistic field $a_{\rm nr}$ satisfies
\be
\nabla_{(r,z)}^2 a_{\rm nr}-\frac{2}{r}\pr_r a_{\rm nr}=-r\mu_0 J\cdot \label{maxwell3}
\ee
We can therefore decompose the total field $a(r,z)$ into the sum of a non-relativistic part $a_{\rm nr}$ (solution of Eq.(\ref{maxwell3})) and a relativistic contribution $a_{\rm rel}$ by setting $a=a_{\rm nr}+a_{\rm rel}$. The relativistic contribution $a_{\rm rel}$ will therefore be a solution of
\bea
\nabla_{(r,z)}^2 a_{\rm rel}-\frac{2}{r}\pr_r a_{\rm rel}&=& -\left\{\left(\pr_r a_{\rm nr}+\pr_r a_{\rm rel}\right)\pr_r \rho\right.+\cdots\non\\
&&\left.\left(\pr_z a_{\rm nr}+\pr_z a_{\rm rel}\right)\pr_z \rho\right\}\label{maxwell4}
\eea
This avoids dealing with point-like sources representing the current loop and the solenoid  in cylindrical coordinates. The source of the field equations now lies in the non-relativistic contribution $a_{\rm nr}(r,z)$.

A current loop of radius $l$ corresponds to a current density located on a infinitely thin ring: $J\sim \delta(z)\cdot\delta(r-l)$ while a solenoid of finite length $L$ and of radius $l$ corresponds to a current density located on an infinitely thin sheet located at $r=l$ and $z\in[-\frac{L}{2},\frac{L}{2}]$ \cite{jackson}. Analytical expressions of the vector potential $A_\varphi$ in both cases can be derived from the Biot-Savart law, expressions that of course  verify Eq.(\ref{maxwell3}). The non-relativistic solution $a_{\rm nr}^{\rm loop}$ for the current loop is given by \cite{jackson,landau}:
\bea
a_{\rm nr}^{\rm loop}(r,z)&=&\frac{\mu_0 I}{2\pi}\sqrt{(l+r)^2+z^2}\times \non\\
&& \left\{\frac{l^2+r^2+z^2}{(l+r)^2+z^2}K(k^2)-E(k^2)\right\}\label{a_loop}
\eea
where $I$ is the steady current carried by the wire, 
$$
k^2=4r l \left((l+r)^2+z^2\right)^{-1}
$$ 
and 
$$
K(k^2)=\int_0^{\pi/2}\left(1-k^2\sin^2(\varphi)\right)^{-1/2}d\varphi,
$$ 
$$
E(k^2)=\int_0^{\pi/2}\left(1-k^2\sin^2(\varphi)\right)^{1/2}d\varphi
$$ 
are the complete elliptic integrals of the first and second kind respectively.\\
For the solenoid of finite length $L$, the non-relativistic solution $a_{\rm nr}^{\rm sol}(r,z)$ is given by \cite{callaghan}:
\bea
a_{\rm nr}^{\rm sol}(r,z)&=&\frac{\mu_0 n I}{4\pi}\sqrt{lr}\times \left[\xi k\left(\frac{k^2+g^2-g^2k^2}{k^2g^2}K(k^2)\right.\right.\non\\
&& \left.\left.-\frac{E(k^2)}{k^2}+\frac{g^2-1}{g^2}\Pi(g^2,k^2)\right)\right]^{\xi_+}_{\xi_-}\label{a_sol}
\eea
where $n$ is the number of wire loops per unit length, 
$$
k^2=\frac{4r l}{ \left((l+r)^2+\xi^2\right)},
$$ 
$$
g^2=\frac{4r l}{(l+r)^{2}}
$$
 $$
 \xi_{\pm}=z\pm \frac{L}{2}
$$ 
and 
$$
\Pi(g^2,k^2)=\int_0^{\pi/2}\left(1-g^2\sin^2(\varphi)\right)^{-1}\left(1-k^2\sin^2(\varphi)\right)^{-1/2}d\varphi
$$
is the complete elliptic integral of the third kind.\\
In order to be used as source terms in Eq.(\ref{maxwell4}), the gradients of $a_{\rm nr}(r,z)$ can be obtained analytically from the formulae Eqs(\ref{a_loop}-\ref{a_sol}) and the properties of complete elliptic functions. 
\begin{widetext}
If we set $r=u l$,
$z= v L$ and $a_{\rm nr,rel}\rightarrow a_{\rm nr,rel}/\left(\mu_0 I l\right)$ ($a_{\rm nr,rel}\rightarrow a_{\rm nr,rel}/\left(\mu_0 n I l L\right)$ for the solenoid),  Eqs. (\ref{rho1},\ref{lam1},\ref{maxwell4}) now reduce to the following
 set of dimensionless equations
\bea
\nabla^2\rho&=&\cI\frac{L^2}{l^2} \frac{e^\rho}{u^2}\left(\left(\pr_u (a_{\rm nr}+a_{\rm rel})\right)^2+\frac{l^2}{L^2}\left(\pr_v  (a_{\rm nr}+a_{\rm rel})\right)^2\right)\label{rho2}\\
\nabla^2\lambda+\frac{l^2}{L^2}\left(\pr_v\rho\right)^2 &=&\cI\frac{L^2}{l^2}\frac{e^\rho}{u^2}\left(\left(\pr_u (a_{\rm nr}+a_{\rm rel})\right)^2-\frac{l^2}{L^2}\left(\pr_v (a_{\rm nr}+a_{\rm rel})\right)^2\right)\label{lam2}\\
\nabla^2 a_{\rm rel}-\frac{2}{u}\pr_u a_{\rm rel}&=&-\left(\pr_u (a_{\rm nr}+a_{\rm rel})\pr_u\rho+\frac{l^2}{L^2}\pr_v (a_{\rm nr}+a_{\rm rel})\pr_v \rho\right) \label{maxwell5}\\
0&=&-\pr_v\lambda+u\pr_u\rho\pr_v\rho-\pr_v\rho+2\cI\frac{e^\rho}{u}\pr_u (a_{\rm nr}+a_{\rm rel})\pr_v (a_{\rm nr}+a_{\rm rel}) \label{Rrz}
\eea
where $\nabla^2=\pr_u^2 +\frac{1}{u}\pr_u+\frac{l^2}{L^2}\pr_v^2$. In the following, we will solve Eqs. (\ref{rho2}-\ref{maxwell5}) numerically and use the last of the Einstein equations Eq.(\ref{Rrz}) for validation check.
\end{widetext}
The dimensionless magneto-gravitational coupling for the current loop and the solenoid are 
given by
\be
\cI^{\rm loop}=\frac{8\pi G}{c^4}\mu_0 I^2 \; ;\; \cI^{\rm sol}=\frac{8\pi G}{c^4}\mu_0 I^2n^2 l^2\cdot
\label{coupling}
\ee 
Hence it is the square of the total current ($I$ for the loop and $I n L$ for the solenoid) that sources the gravitational field\footnote{The magneto-gravitational coupling $\cI$ can also be rewritten:
$$
\cI=8\pi \left(\frac{I}{I_{\rm Pl}}\right)^2
$$
where $I$ represents the total current involved and 
$I_{\rm Pl}=c^2/\sqrt{G \mu_0}=9.8169\times 10^{24}A$ 
is the Planck current.
}

\subsection{Boundary conditions}
Solving the system (\ref{rho2}-\ref{maxwell5}) requires the specification of boundary conditions. On the axis of symmetry, $r=0$, space-time must be smooth
such that we have $\pr_r \rho|_{r=0}=\pr_r \lambda|_{r=0}=\pr_r a|_{r=0}=0$. Far away from the current loop or the solenoid ($(u,v)\rightarrow (+\infty,\pm \infty)$), these devices behave as magnetic dipoles and the space-time is asymptotically flat (see also \cite{bonnor54}). The magnetic component $a$ is then ruled by Eq.(\ref{maxwell2}) with $\pr_u a \pr_u\rho+\pr_v a \pr_v \rho\approx 0$ (i.e., the non-relativistic Maxwell equation). This condition is achieved if $a_{\rm rel}$ vanishes and $\pr_v a_{\rm nr}=u\pr_u \psi$ and $\pr_u a_{\rm nr}=-u\pr_v\psi$ with $\psi$ an harmonic function called the scalar magnetic potential \cite{bonnor54,bonnor60}. For magnetic dipoles, the (dimensionless) scalar magnetic potential is given by
\be
\psi=\frac{v}{4}\left(u^2+\frac{L^2}{l^2}v^2\right)^{-3/2}\cdot
\ee
The metric functions $\rho$ and $\lambda$ are therefore ruled by the following equations at large distances ($(u,v)\rightarrow (+\infty,\pm \infty)$, $\rho\ll 1$ so that $e^\rho\approx 1$):
\bea
\nabla^2\rho&=&\cI\frac{L^4}{l^4}\left(\left(\pr_u \psi\right)^2+\frac{l^2}{L^2}\left(\pr_v \psi\right)^2\right)\non\\
\nabla^2\lambda&=&\cI\frac{L^4}{l^4}\left(\frac{l^2}{L^2}\left(\pr_v \psi\right)^2-\left(\pr_u \psi\right)^2\right)\cdot\non
\eea
 
The asymptotic behaviors for the metric fields around the current loop and the solenoid are therefore given by
\bea
\rho&\sim&\frac{\mathcal{C_I}}{32}\frac{L^4}{l^4}v^2\left(u^2+\frac{L^2}{l^2}v^2\right)^{-3}\label{asymp}\\
\lambda&\sim&\frac{\mathcal{C_I}}{16}\frac{L^2}{l^2}\left[2u^2\left(u^2+\frac{L^2}{l^2}v^2\right)^{-3}\right.\non\\
&&-\frac{L^2}{l^2}\frac{v^2}{2}\left(u^2+\frac{L^2}{l^2}v^2\right)^{-3}\non\\
&&\left.-\frac{9u^4}{4}\left(u^2+\frac{L^2}{l^2}v^2\right)^{-4}\right]\cdot\label{asymp2}
\eea
\section{Numerical resolution of field equations}
\subsection{Numerical Method}
We solve Eqs.(\ref{rho2}-\ref{maxwell5}) numerically by using a combination of relaxation and spectral methods. We first introduce a sequence of functions $\rho^{(n)}(u,v)$,
$\lambda^{(n)}(u,v)$ and $a_{\rm rel}^{(n)}(u,v)$ for the relaxation algorithm such that
 Eqs.(\ref{rho2}-\ref{maxwell5}) can be approximated by a set of $linear$ inhomogeneous elliptic equations:
 \bea
\nabla^2\rho^{(n+1)}&=& S_1\left[\rho^{(n)},a_{\rm rel}^{(n)}\right]\label{rel1}\\
\nabla^2\lambda^{(n+1)} &=&S_2\left[\rho^{(n)},a_{\rm rel}^{(n)}\right]\\
\nabla^2 a_{\rm rel}^{(n+1)}-\frac{2}{u}\pr_u a_{\rm rel}^{(n+1)}&=&S_3\left[\rho^{(n)},a_{\rm rel}^{(n)}\right]\label{rel3}
\eea
 where the source terms $S_i$ gather all non-linear terms of Eqs.(\ref{rho2}-\ref{maxwell5})
 but evaluated with the previous state $n$ of the relaxation algorithm. The algorithm starts
 with a state $n=0$ that corresponds to the non-relativistic solution: 
 $\rho^{(0)}(u,v)=\lambda^{(0)}(u,v)=a_{\rm rel}^{(0)}(u,v)=0$. At each relaxation step,
 we solve Eqs.(\ref{rel1}-\ref{rel3}) with a spectral method to compute 
 $\rho^{(n+1)}(u,v)$,
$\lambda^{(n+1)}(u,v)$ and $a_{\rm rel}^{(n+1)}(u,v)$ before iterating. The algorithm is stopped when the relative  update of the three fields , averaged over the spatial domain, reaches some tolerance threshold.
 
 To solve Eqs.(\ref{rel1}-\ref{rel3}) at each relaxation step, we develop each field $f(u,v)$ and each source term $S_i(u,v)$ as a truncated Fourier series in the $v-$direction:
 $$
 f(u,v)=\sum_{k=0}^N \hat{f}_k(u) \cos\left(\frac{k\pi}{V}v\right)\; ;\; (u,v)\in\left[0,U\right]\times\left[-V,+V\right]
 $$
since all the fields are even functions of $v$ due to cylindrical symmetry and where
$$
\hat{f}_k(u) =\frac{1}{V}\int_{-V}^V f(u,v)\cos\left(\frac{k\pi}{V}v\right)dv\cdot
$$
In Fourier space, Eqs.(\ref{rel1}-\ref{rel3}) now become a set of linear inhomogeneous ODEs
 \bea
%\frac{d^2\hat{\rho}^{(n+1)}_k(u)}{du^2}+\frac{1}{u}\frac{d\hat{\rho}^{(n+1)}_k(u)}{du}-\left(\frac{k\pi}{V}\right)^2&=& \hat{S}_{k,1}(u)\label{rel4}\\
%\frac{d^2\hat{\lambda}^{(n+1)}_k(u)}{du^2}+\frac{1}{u}\frac{d\hat{\lambda}^{(n+1)}_k(u)}{du}-\left(\frac{k\pi}{V}\right)^2&=& \hat{S}_{k,2}(u)\\
%\frac{d^2\hat{a_{\rm rel}}^{(n+1)}_k(u)}{du^2}-\frac{1}{u}\frac{d\hat{a_{\rm rel}}^{(n+1)}_k(u)}{du}-\left(\frac{k\pi}{V}\right)^2&=& \hat{S}_{k,3}(u)\label{rel6}
\frac{d^2\hat{\rho}_k}{du^2}+\frac{1}{u}\frac{d\hat{\rho}_k}{du}-\frac{l^2}{L^2}\left(\frac{k\pi}{V}\right)^2\hat{\rho}_k&=& \hat{S}_{k,1}(u)\label{rel4}\\
\frac{d^2\hat{\lambda}_k}{du^2}+\frac{1}{u}\frac{d\hat{\lambda}_k}{du}-\frac{l^2}{L^2}\left(\frac{k\pi}{V}\right)^2\hat{\lambda}_k&=& \hat{S}_{k,2}(u)\\
\frac{d^2\hat{a_{\rm rel}}_k}{du^2}-\frac{1}{u}\frac{d\hat{a_{\rm rel}}_k}{du}-\frac{l^2}{L^2}\left(\frac{k\pi}{V}\right)^2\hat{a_{\rm rel}}_k&=& \hat{S}_{k,3}(u)\label{rel6}
\eea
for $k=1,\cdots,N$ and where we have omitted the relaxation index $n+1$. Eqs.(\ref{rel4}-\ref{rel6}) can be solved
as a boundary value problem whose boundary conditions at $u=0$ and $u=U\gg 1$ are the Fourier transforms of the conditions derived in section II.B. In the present paper, we have used the algorithm described in \cite{bvp} for the numerical resolution of Eqs.(\ref{rel4}-\ref{rel6}) for each Fourier mode $k$ at
each relaxation step $n$.

\begin{figure}[ht!]
\includegraphics[trim={0 0 5cm 12cm},clip=true,scale=0.4]{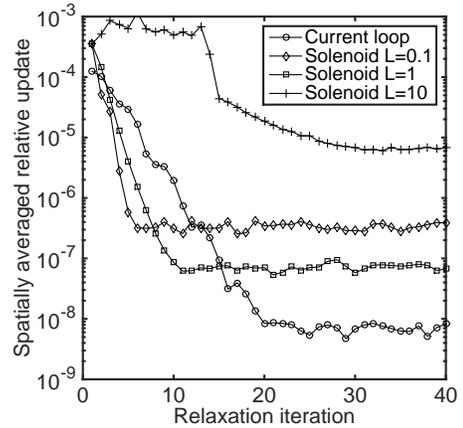}
\caption{Relative update on the fields Eq.(\ref{update}), averaged over the spatial domain,
as a function of the relaxation iteration $n$}\label{fig1}
\end{figure}

Figure \ref{fig1} represents the relative update on the fields 
\be
\frac{1}{3}\left\{\Big|\frac{\rho^{(n+1)}}{\rho^{(n)}}-1\Big|+
\Big|\frac{\lambda^{(n+1)}}{\lambda^{(n)}}-1\Big|+\Big|\frac{a_{\rm rel}^{(n+1)}}{a_{\rm rel}^{(n)}}-1\Big|\right\}\label{update}
\ee
averaged over the spatial domain $\left[0,U\right]\times\left[-V,+V\right]$, as a function of the relaxation step $n$. In the relaxation algorithm, the relative updates on the fields decrease exponentially with the number of iterations before settling to some plateau. The algorithm described above therefore converges toward the solution of Eqs.(\ref{rho2}-\ref{maxwell5}).

%%%% FIG 1 %%%%%%
%%%% FIG 2 %%%%%%
\begin{figure}[ht!]
\includegraphics[trim={4cm 8cm 3cm 8cm},clip=true,scale=0.4]{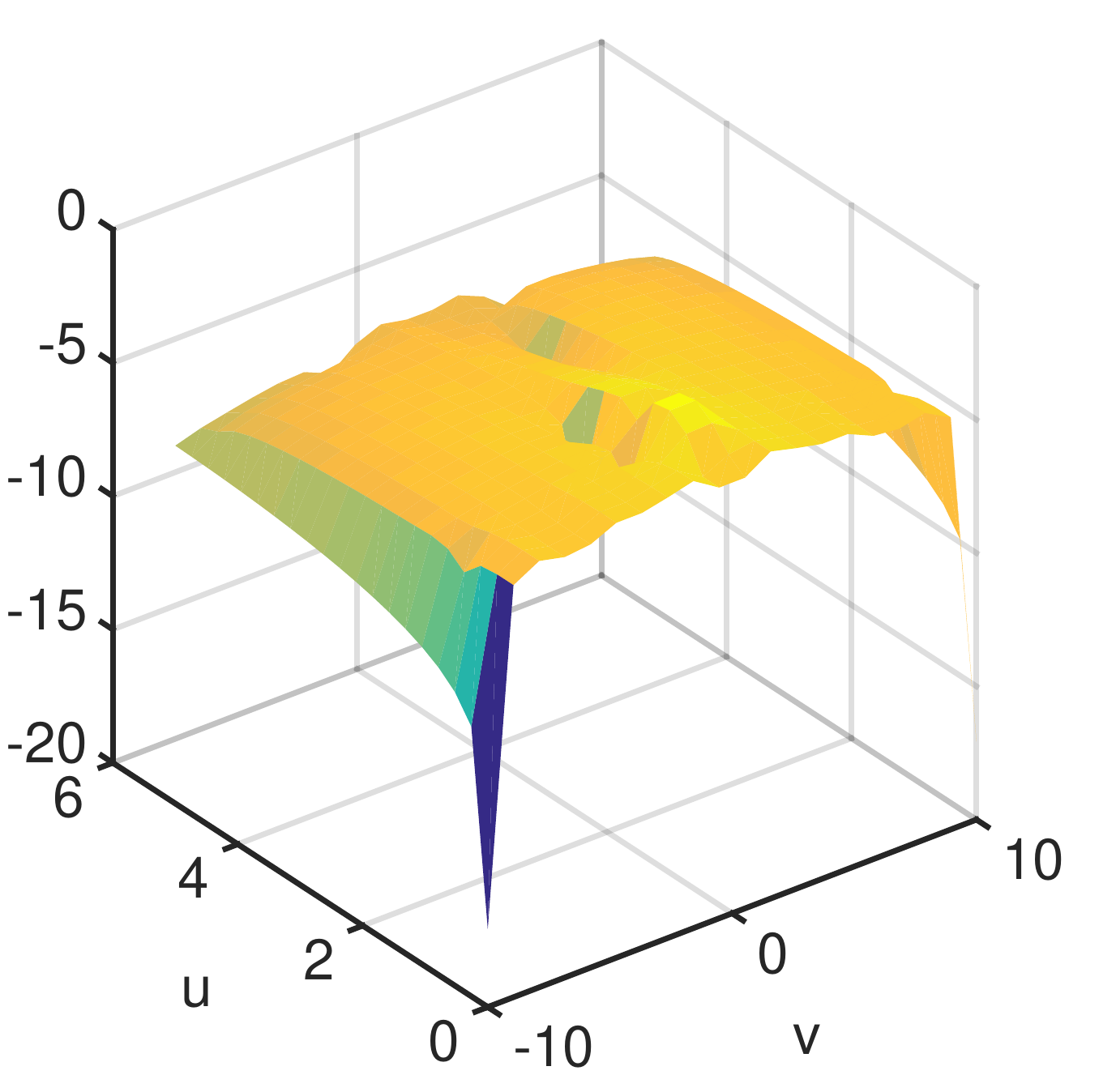}
\caption{
Right hand side of Eq.(\ref{Rrz})  (in logarithmic scale) at the end of the relaxation algorithm (case of a solenoid with $L=0.1$ and $\cI=1$)}\label{fig2}
\end{figure}
%% in response to referee's comments
We can also have a glimpse on the numerical accuracy of the solution by evaluating the right hand side of the off-diagonal Einstein equation, Eq.(\ref{Rrz}), with the fields obtained at the end of the relaxation algorithm. This is illustrated in Figure \ref{fig2} in logarithmic scale, the rms of the residual is about $3\times 10^{-5}$ for field values of order unity ($\cI=1$). The numerical error is more important around the sources of current and are mostly dominated by rounding errors in the evaluation of the elliptic functions that compose the magnetic potential $a_{\rm nr}$ of the current loop and the solenoid\footnote{This can be viewed by evaluating the classical Maxwell equation Eq.(8) with the numerical implementation of $a_{\rm nr}$.}.\\
\\
We can now present the space-times curved by the magnetic fields of the current loop and the solenoid.
\subsection{Numerical Results}
\begin{figure*}[ht!]
\begin{tabular}{ccc}
\includegraphics[trim={2.5cm 0 5cm 5cm},clip=true,scale=0.4]{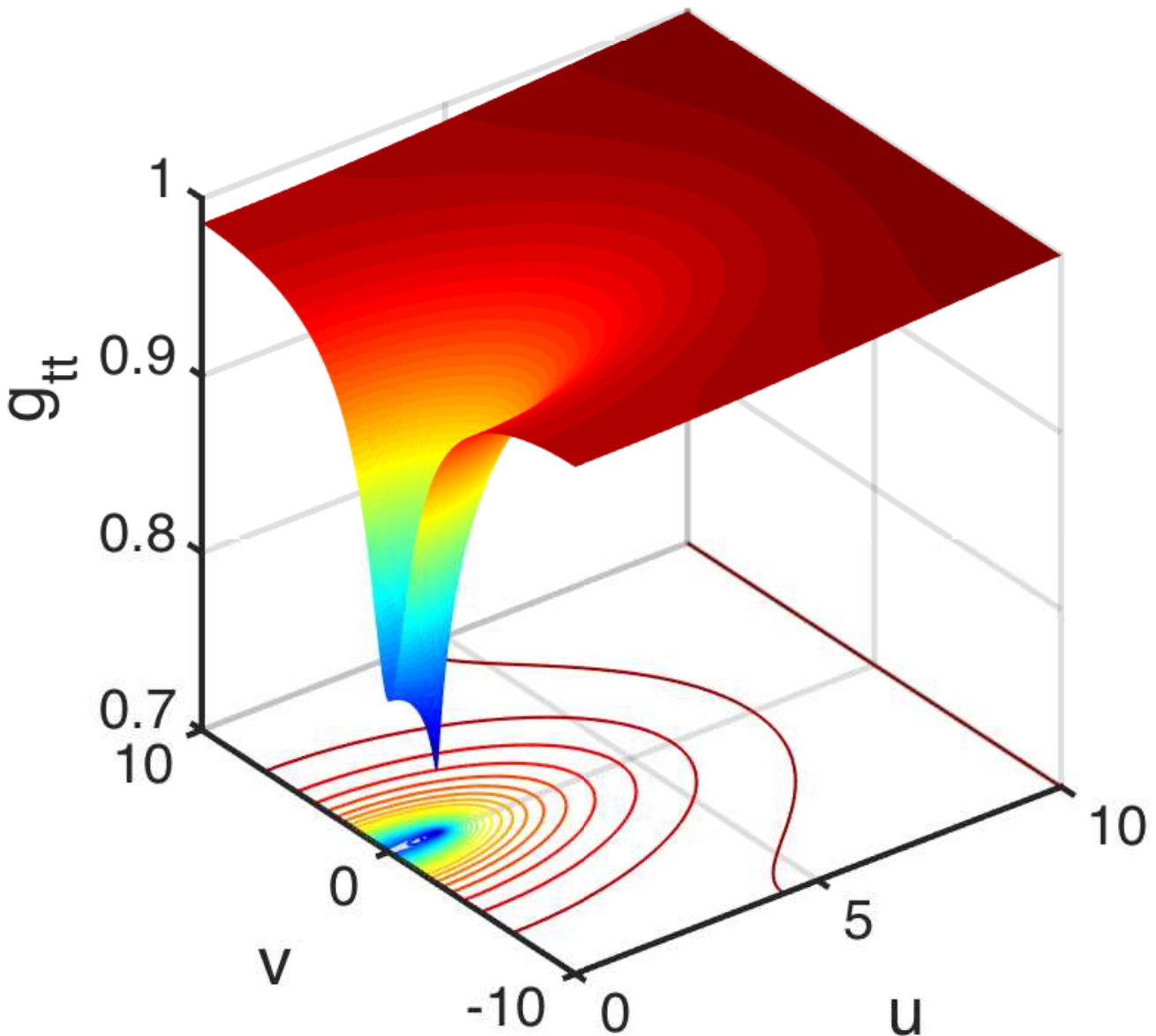} &
\includegraphics[trim={2.5cm 0 5cm 5cm},clip=true,scale=0.45]{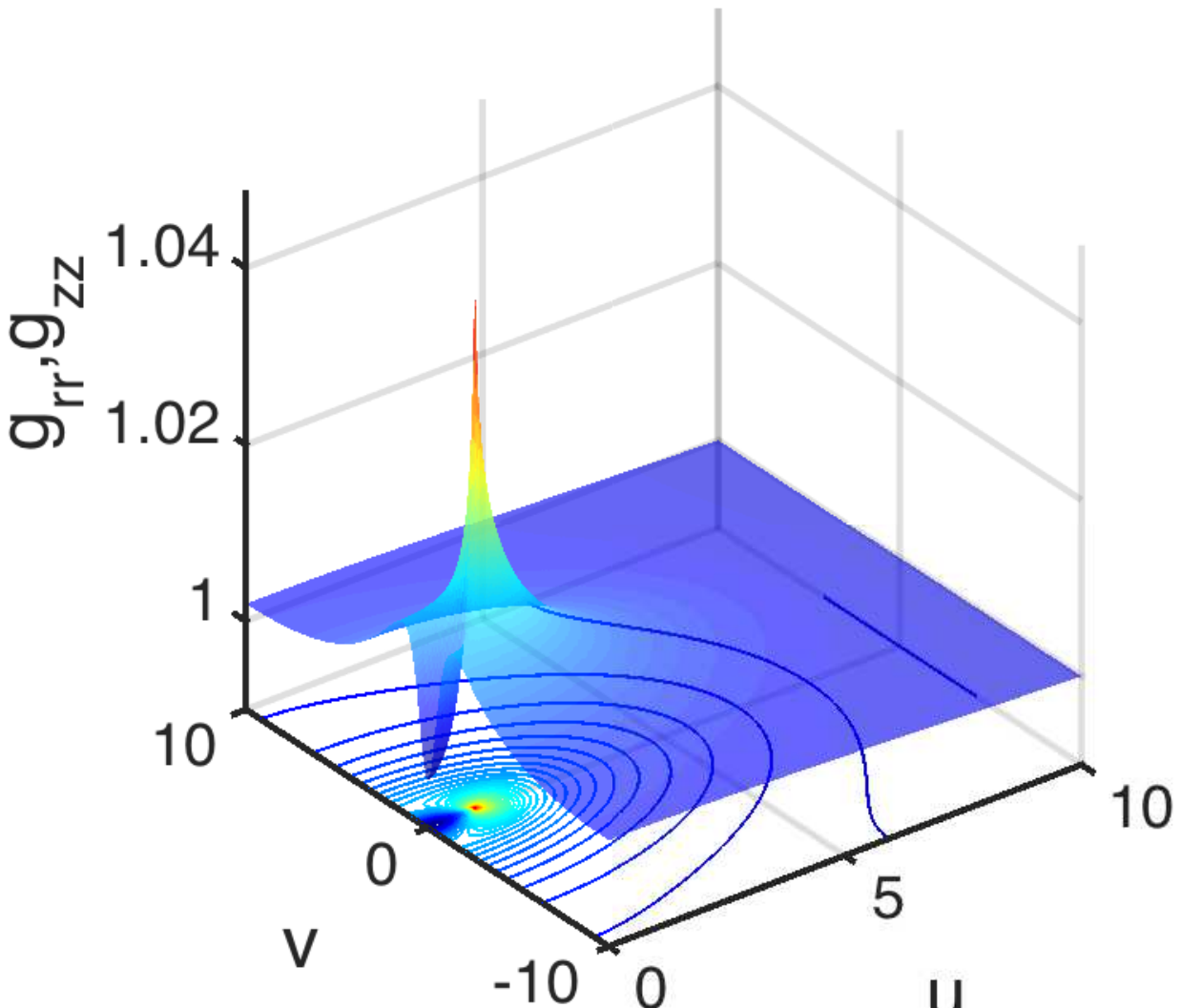} &
\includegraphics[trim={2.5cm 0 5cm 5cm},clip=true,scale=0.45]{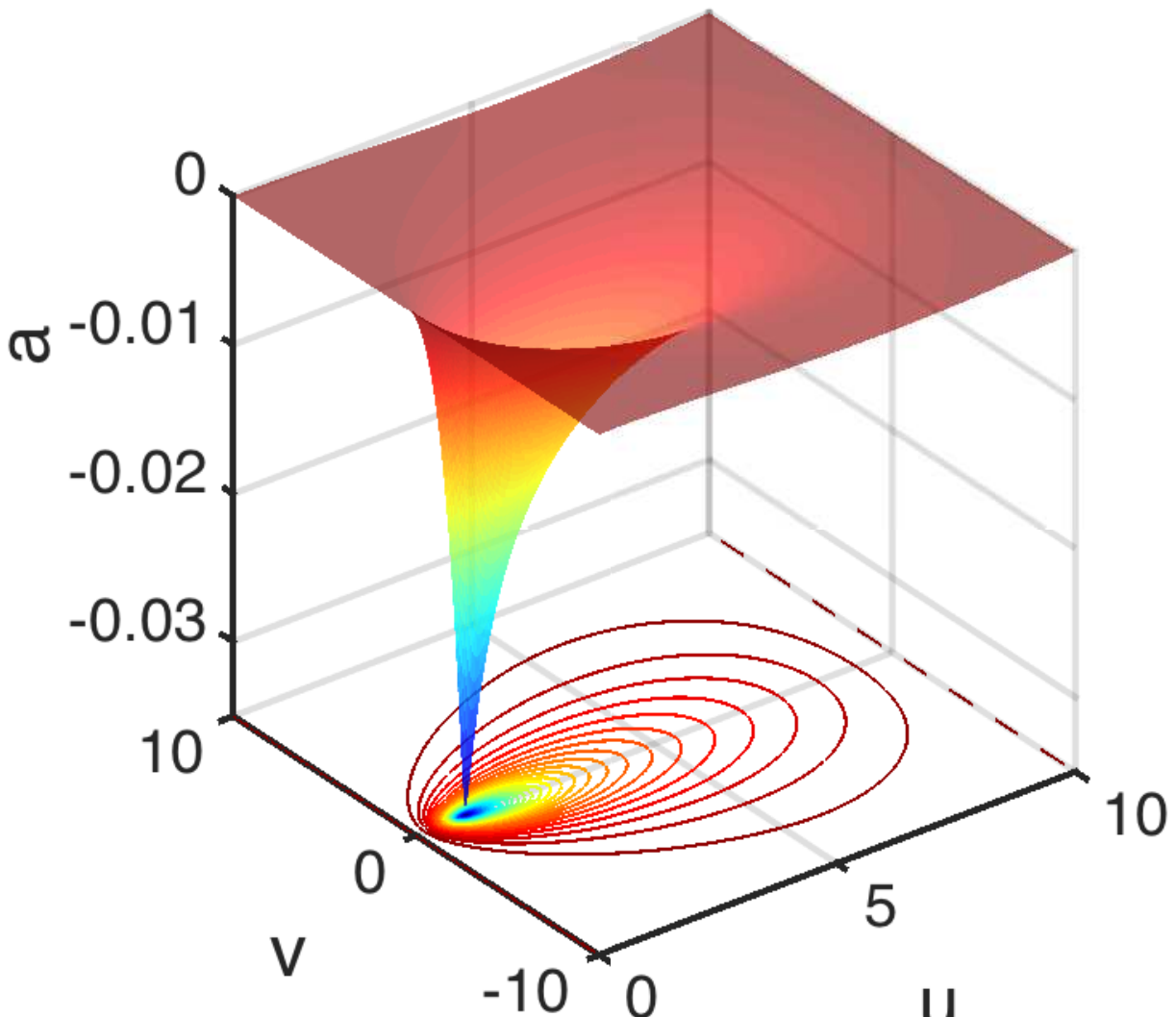}
\end{tabular}  
\caption{Metric components and relativistic part of the magnetic potential for the current loop for $\cI=1$ (left panel: $g_{tt}=-r^2g^{\varphi\varphi}=\exp(\rho)$, central panel: 
$g_{rr}=g_{zz}=\exp(\lambda)$, right panel: $a_{\rm rel}$) }
    \label{fig3}
\end{figure*}

\begin{figure*}[ht!]
\begin{tabular}{ccc}
\includegraphics[trim={1cm 0 3cm 0},clip,scale=0.3]{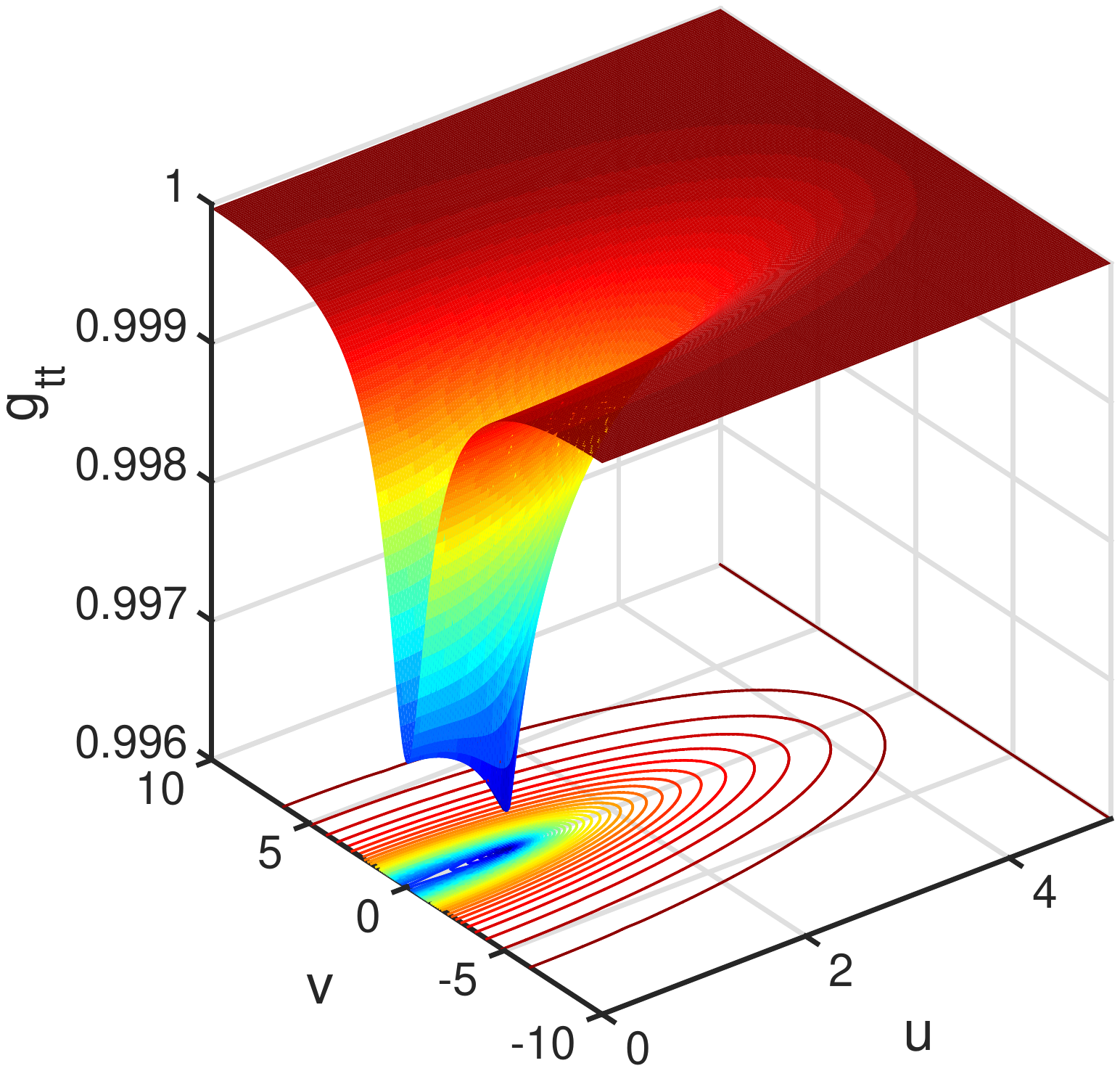} &
\includegraphics[trim={1cm 0 3cm 0},clip,scale=0.3]{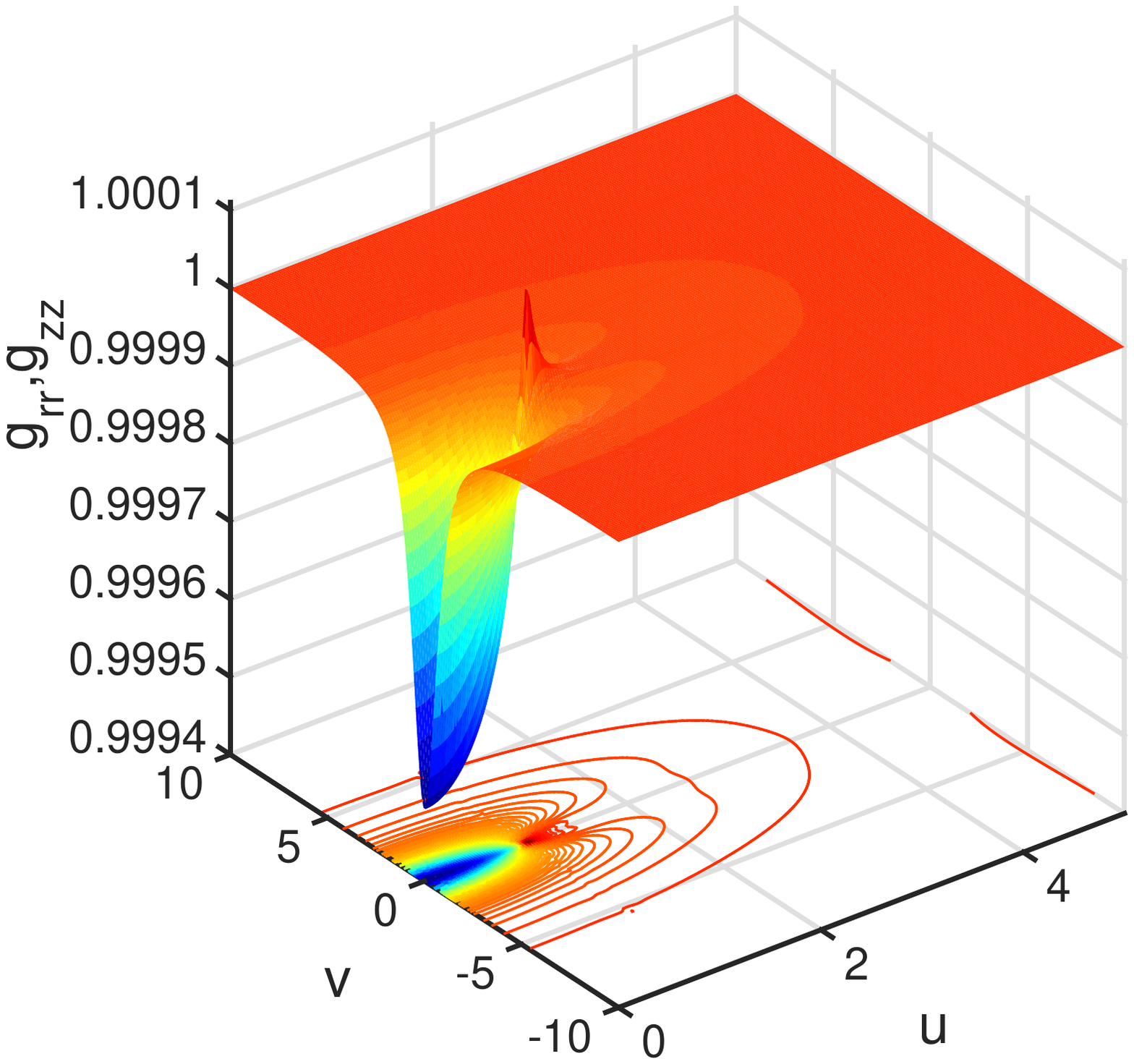} &
\includegraphics[trim={1cm 0 3cm 0},clip,scale=0.3]{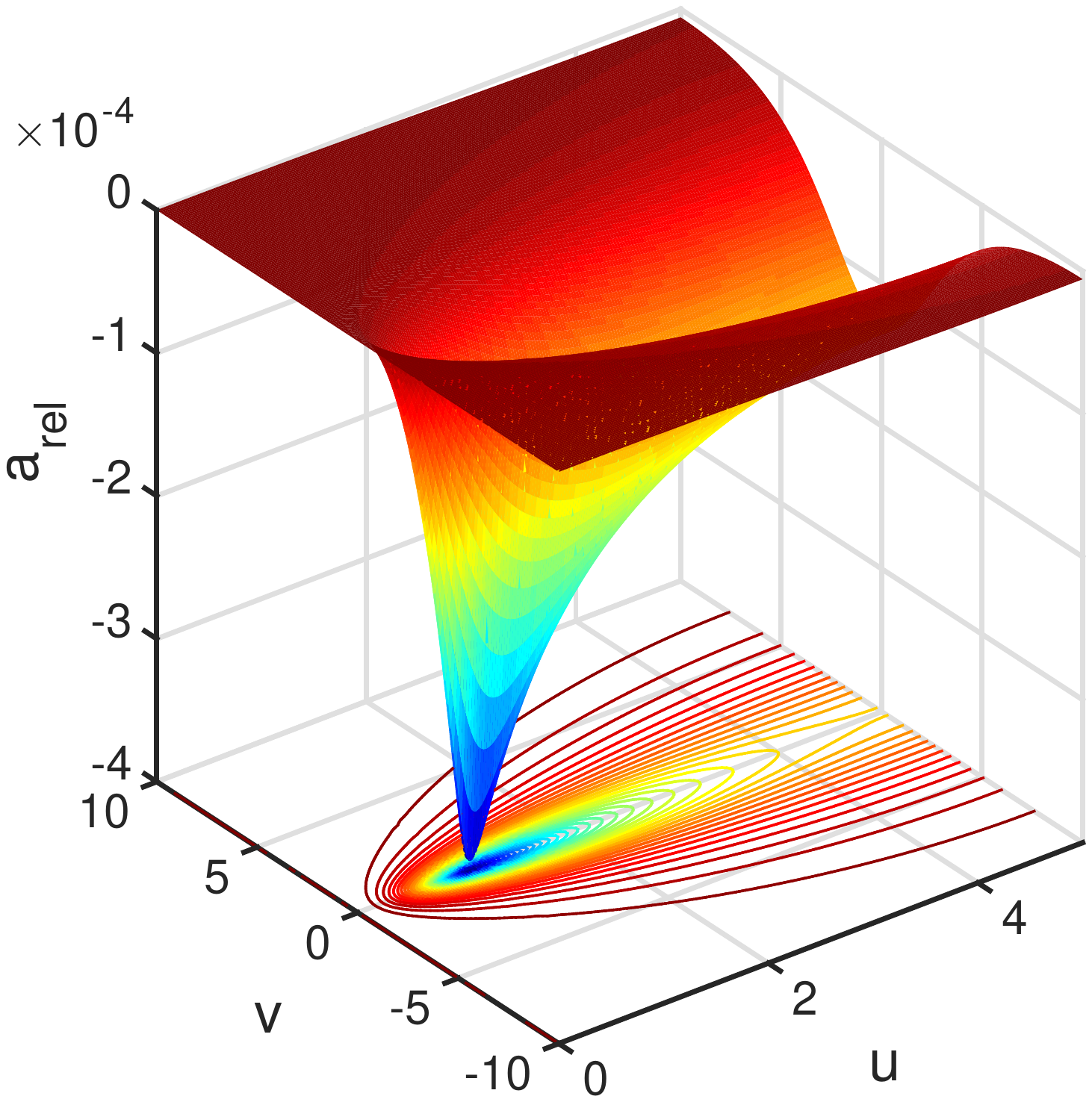}\\
\includegraphics[trim={1cm 0 3cm 10cm},clip,scale=0.3]{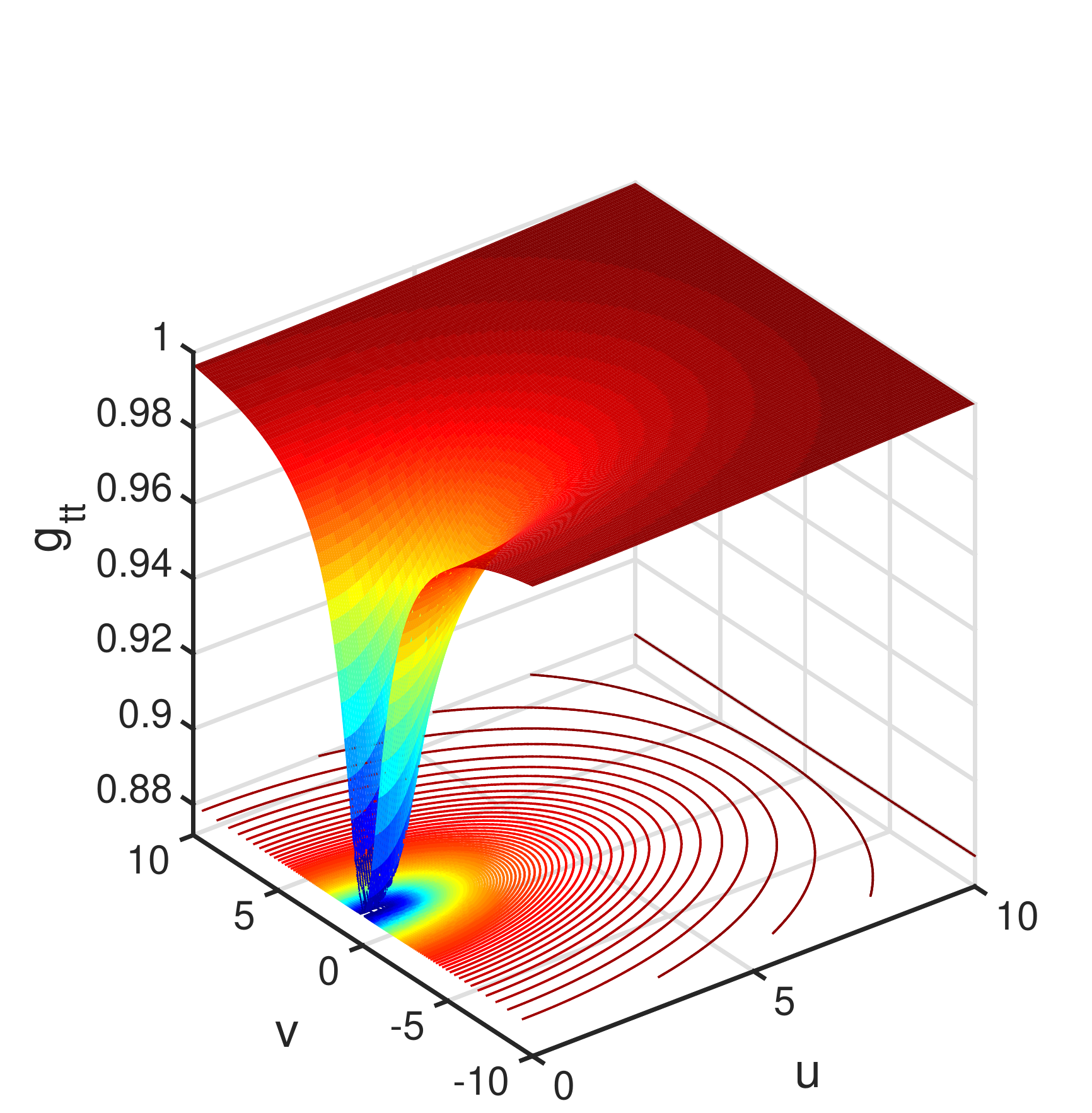} &
\includegraphics[trim={1.cm 0 3cm 10cm},clip,scale=0.3]{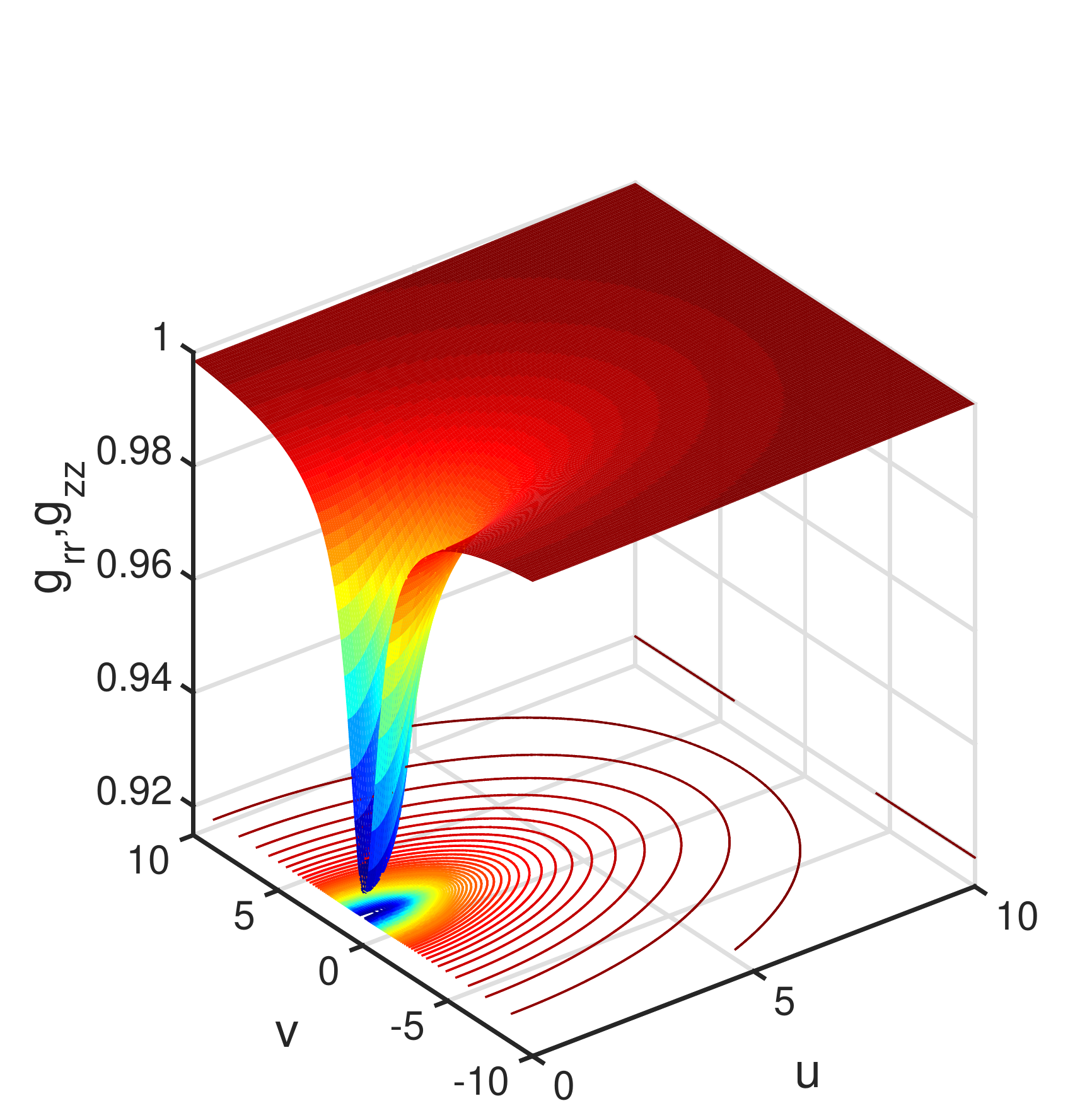} &
\includegraphics[trim={1.cm 0 3cm 10cm},clip,scale=0.3]{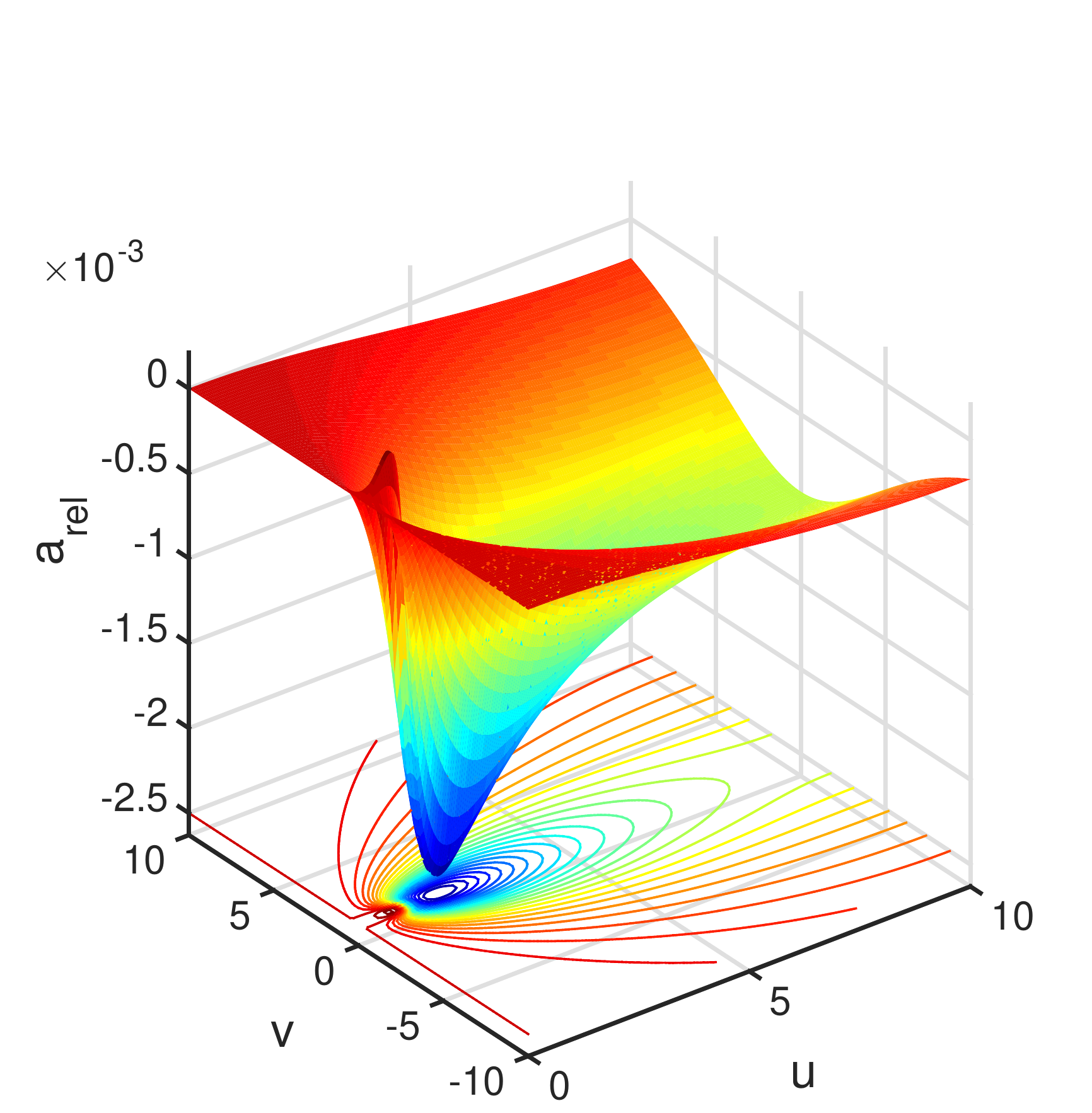}\\
\includegraphics[trim={1.cm 0 3cm 10cm},clip,scale=0.3]{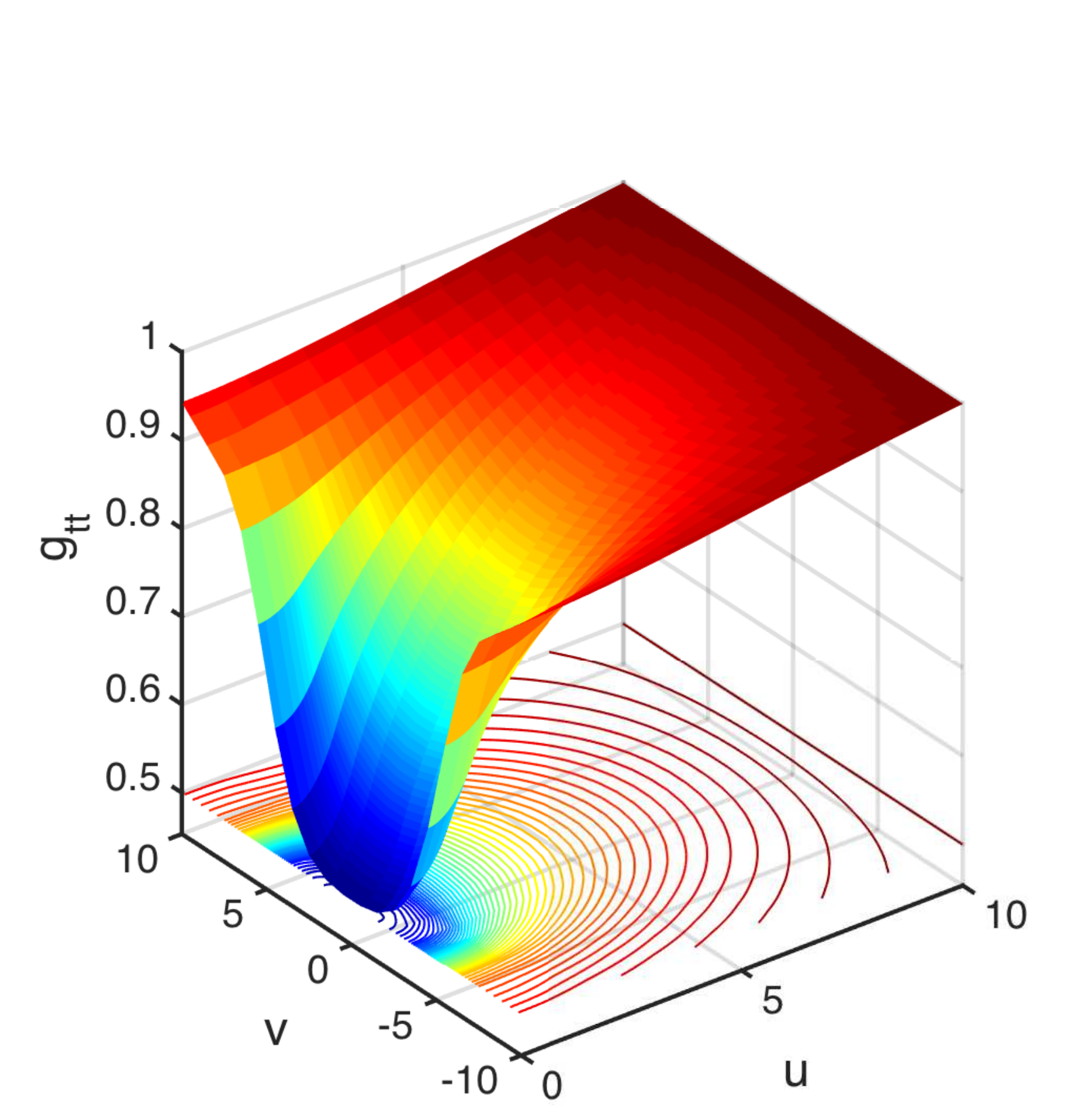} &
\includegraphics[trim={1.cm 0 3cm 10cm},clip,scale=0.3]{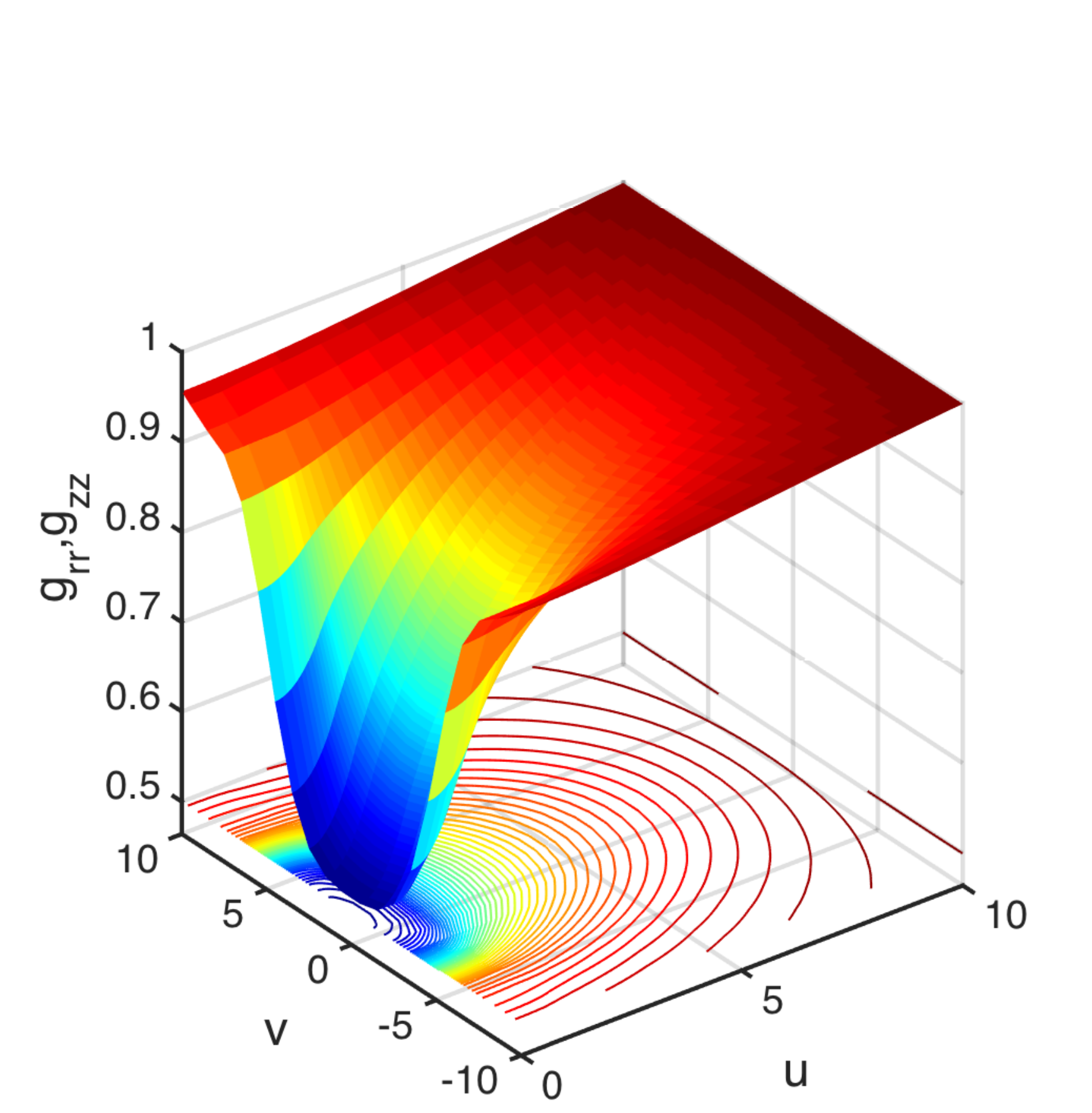} &
\includegraphics[trim={1.cm 0 3cm 10cm},clip,scale=0.3]{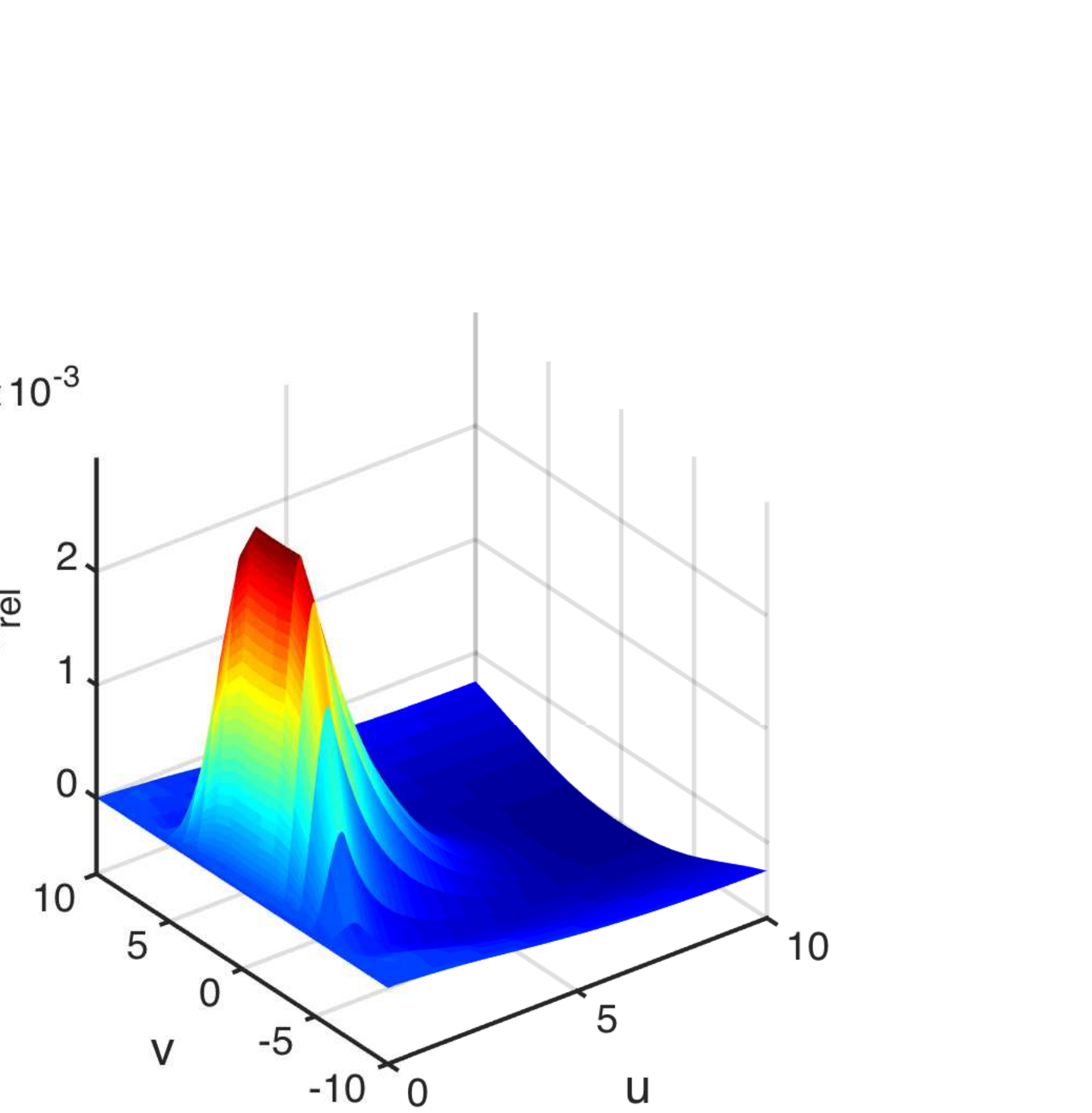}
\end{tabular}  
\caption{Metric components and relativistic part of the magnetic potential for solenoids of different lengths for $\cI=1$ (left panel: $g_{tt}=-r^2g^{\varphi\varphi}=\exp(\rho)$, central panel: 
$g_{rr}=g_{zz}=\exp(\lambda)$, right panel: $a_{\rm rel}$ ; first row: $L=0.1$, second row: $L=1$, third row: $L=10$) }
    \label{fig4}
\end{figure*}

Figures \ref{fig3} and \ref{fig4} present the metric functions $\rho$ and $\lambda$ as well as the relativistic part of the magnetic potential $a_{\rm rel}$ for the current loop (Figure \ref{fig3}) and solenoids of different lengths (Figure \ref{fig4}).
These plots have been obtained from the numerical resolution of Eqs.(\ref{rho2}-\ref{maxwell5}) with the boundary conditions Eqs.(\ref{asymp}) using the numerical method presented in the previous section. 
%%%%% FIG 2 %%%%%%
%%%% FIG 3 %%%%%%

Metric functions exhibit a peak at the loop location $(u=1,v=0)$ in Figure \ref{fig3} and
we can notice the similarity between the solutions of the current loops and the solenoid of smaller length (Figure \ref{fig4}, first row: $L=0.1$). Space-time is mostly curved along the radial direction ($v=0$) inside the loop ($u<1$) and becomes quickly flat outside of the loop. 

When the solenoid length is increased (Figure \ref{fig4} from top to bottom),  we can see how the metric potential wells deepens and widens along the central $z-$axis. With longer solenoids, the total electric current involved is increased and so does the space-time deformation at the origin of coordinates. The peak in $\lambda$ near the solenoid location is also smoothed when the solenoid length is increasing (Figure \ref{fig4}, central panels from top to bottom). We also have that space-time deformation occurs mostly inside the solenoid and space-time becomes quickly flat outside the solenoid.

\begin{figure}[ht!]
\includegraphics[trim={1.cm 0 4cm 10cm},clip=true,scale=0.4]{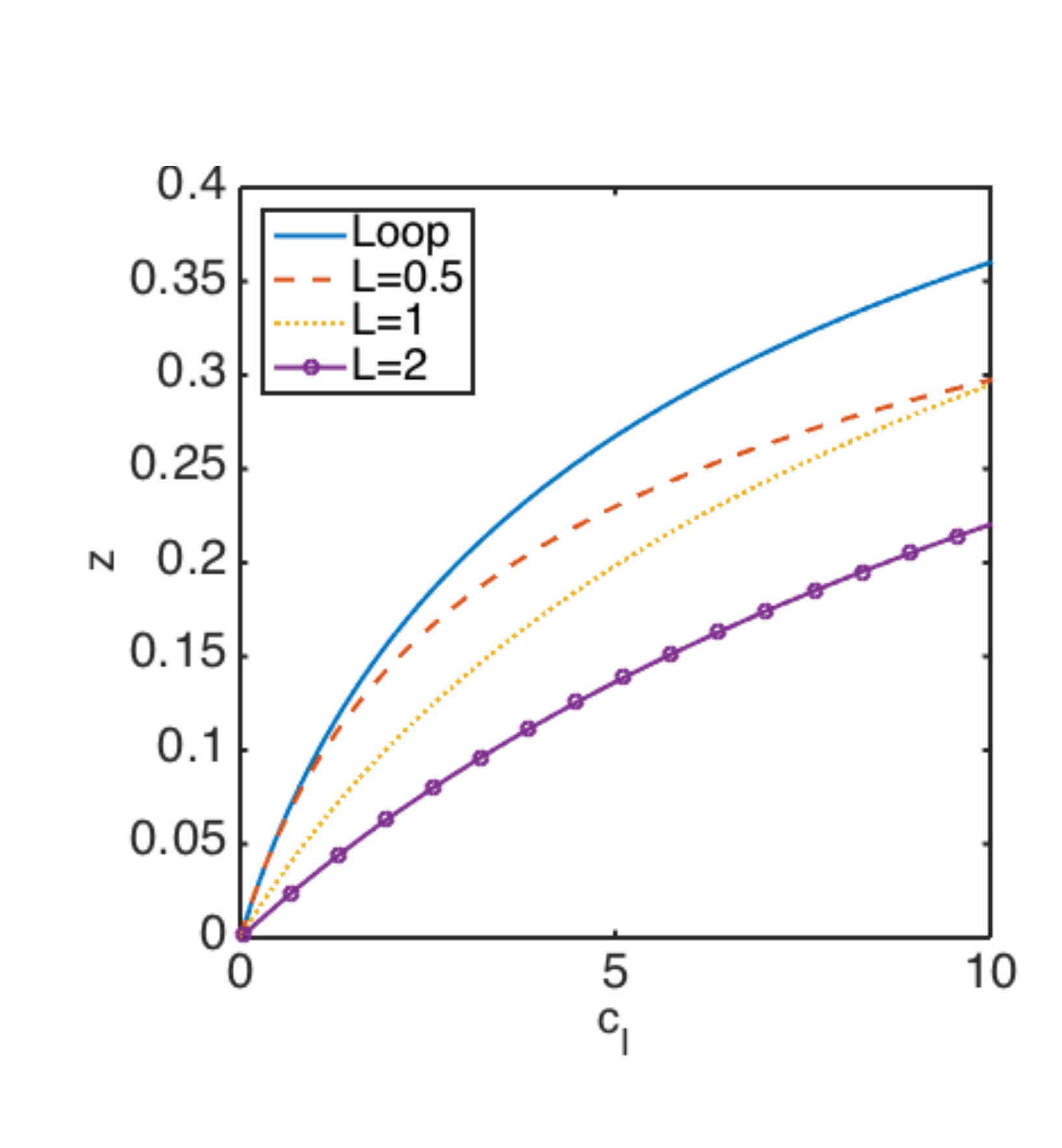}
\caption{Gravitational redshift of a light source located at the center of coordinates and a receiver at spatial infinity, as a function of the magneto-gravitational coupling $\cI$}\label{fig5}
\end{figure}

The total magnetic potential $a=a_{\rm nr}+a_{\rm rel}$ is smaller than in the classical case since we have $a_{\rm rel}<0$ on average (Figures \ref{fig3} and \ref{fig4}, right panels). In general relativity, the electric current produces both magnetic and gravitational fields, leaving less energy to the former than it does in electromagnetism on flat space-times.
\\
These metric potential wells will produce light deflexion as well as gravitational redshift which will be maximal for a light source located at the origin of coordinates. For an observer located at spatial infinity (where $g_{tt}\rightarrow 1$) and a source located at the origin of coordinates, the gravitational redshift is simply $z=1-\exp(-\rho(0,0)/2)$, and is of the order of magnitude of the magneto-gravitational coupling $\cI$ (see also Fig. \ref{fig5}). The precision achieved by optical lattice clocks in the measurement of a transition frequency is of the order $10^{-15}$ \cite{blatt}. Achieving such a gravitational redshift with single-layered solenoids would require $\cI\approx 10^{-15}$, i.e. for an electric current
of $1kA$, $n=100$ it would require a solenoid length of about $10^{11}m$. Gravitational redshift therefore does not seem to be appropriate to detect gravitational fields artificially generated by coils with current superconducting technology. As we shall see further, Michelson interferometers with Fabry-P\'erot cavities will be more appropriate to attempt such detection of artificially generated gravitational fields.

Fig. \ref{fig5} illustrates how the gravitational redshift $z$ evolves with the magneto-gravitational coupling $\cI$ Eq.(\ref{coupling}). For $\cI\ll 1$, the redshift $z$ varies linearly with $\cI$ while a space-time singularity appears at the center of coordinates, $\rho(0,0)\rightarrow -\infty$ ($z\rightarrow 1$) when $\cI\rightarrow\infty$.

We can now focus on the relativistic effect of the deviation of light by an electric current, through the induced deformation of space-time by magnetic fields. This will be treated of the next section.

\section{The bending of light by magnetic fields in general relativity}
In this section, we establish the general pattern of light geodesics in strongly curved space-times through numerical integration techniques. The geodesic curves $x^\mu(s)$ of neutral particles are solutions of 
 $$
 \frac{d^2 x^\alpha}{ds^2}+\Gamma^{\alpha}_{\beta\gamma}\frac{dx^\beta}{ds}\frac{dx^\gamma}{ds}=0
 $$
 where $s$ is some affine parameter. Introducing the metric ansatz Eq.(\ref{weyl}), this gives the following set of ordinary differential equations:
 \begin{widetext}
 \bea
 \frac{d^2 t}{ds^2}&+&\frac{\pr \rho}{\pr r}\frac{dt}{ds}\frac{dr}{ds}+\frac{\pr \rho}{\pr z}\frac{dt}{ds}\frac{dz}{ds}=0\label{geo1}\\
 \frac{d^2 z}{ds^2}&+&\frac{c^2}{2}e^{\rho-\lambda}\frac{\pr \rho}{\pr z}\left(\frac{dt}{ds}\right)^2+\frac{1}{2}\frac{\pr \lambda}{\pr z}\left(\frac{dz}{ds}\right)^2+\frac{r^2}{2}e^{-\rho-\lambda}\frac{\pr \rho}{\pr z}\left(\frac{d\varphi}{ds}\right)^2+\frac{\pr \lambda}{\pr r}\frac{dz}{ds} \frac{dr}{ds}-\frac{1}{2}\frac{\pr \lambda}{\pr z}\left(\frac{dr}{ds}\right)^2=0\label{geo2}\\
 \frac{d^2 r}{ds^2}&+&\frac{c^2}{2}e^{\rho-\lambda}\frac{\pr \rho}{\pr r}\left(\frac{dt}{ds}\right)^2-\frac{1}{2}\frac{\pr \lambda}{\pr r}\left(\frac{dz}{ds}\right)^2+\frac{r}{2}e^{-\rho-\lambda}\left(r\frac{\pr \rho}{\pr r}-2\right)\left(\frac{d\varphi}{ds}\right)^2+\frac{\pr \lambda}{\pr z}\frac{dz}{ds} \frac{dr}{ds}+\frac{1}{2}\frac{\pr \lambda}{\pr r}\left(\frac{dr}{ds}\right)^2=0\label{geo3}\\
 \frac{d^2 \varphi}{ds^2}&+&\left(\frac{2}{r}-\frac{\pr \rho}{\pr r}\right)\frac{dr}{ds}\frac{d\varphi}{ds}-\frac{\pr \rho}{\pr z}\frac{d\varphi}{ds}\frac{dz}{ds}=0\label{geo4}
 \eea
\end{widetext}
Eqs.(\ref{geo1}) and (\ref{geo4}) can be directly integrated to give
\bea
\frac{dt}{ds}&=&\frac{1}{c}e^{-\rho}\\
\frac{d\varphi}{ds}&=&\frac{C}{r^2}e^\rho
\eea
where we chose one integration constant such that the affine parameter $s$ can be identified with the coordinate time $ct$ at spatial infinity (where $\rho\rightarrow 0$)
and where the constant $C$ is related to the angular momentum of the neutral particle. 

For null geodecics, we have that the tangent vector is light-like all along the geodesic curves $g_{\mu\nu}\frac{dx^\mu}{ds}\frac{dx^\nu}{ds}=0$ so that 
\be
e^{-\rho}-\frac{C^2}{r^2}e^\rho-e^{\lambda}\left[\left(\frac{dz}{ds}\right)^2+\left(\frac{dr}{ds}\right)^2\right]=0\cdot\label{lagrangian}
\ee
Putting this constraint into the remaining two geodesic equations Eqs.(\ref{geo2}) and (\ref{geo3}) gives
\begin{widetext}
\bea
\frac{d^2 z}{ds^2}&+&\frac{C^2}{r^2}e^{\rho-\lambda}\frac{\pr \rho}{\pr z}+\frac{1}{2}\left(\frac{\pr \rho}{\pr z}+\frac{\pr \lambda}{\pr z}\right)\left(\frac{dz}{ds}\right)^2+\frac{1}{2}\left(\frac{\pr \rho}{\pr z}-\frac{\pr \lambda}{\pr z}\right)\left(\frac{dr}{ds}\right)^2+\frac{\pr \lambda}{\pr r}\frac{dz}{ds} \frac{dr}{ds}=0\label{geo5}\\
 \frac{d^2 r}{ds^2}&+&\frac{C^2}{r^2}e^{\rho-\lambda}\left(\frac{\pr \rho}{\pr r}-\frac{1}{r}\right) +\frac{1}{2}\left(\frac{\pr \rho}{\pr r}-\frac{\pr \lambda}{\pr r}\right)\left(\frac{dz}{ds}\right)^2
 +\frac{1}{2}\left(\frac{\pr \rho}{\pr r}+\frac{\pr \lambda}{\pr r}\right)\left(\frac{dr}{ds}\right)^2+\frac{\pr \lambda}{\pr z}\frac{dz}{ds} \frac{dr}{ds}=0\label{geo6}
\eea

In the following, we restrict ourselves to planar trajectories in the $(r,z)-$plane by setting
$\varphi=\rm cst$ and $C=0$ (such that $\frac{d\varphi}{ds}=0$). If we now set $s=Sl$ (and $r=ul$, $z=vL$) we finally obtain the following set of dimensionless ODEs:
\bea
\frac{d^2v}{dS^2}&+&\frac{1}{2}\left(\frac{\pr \rho}{\pr v}+\frac{\pr \lambda}{\pr v}\right)\left(\frac{dv}{dS}\right)^2+\frac{l^2}{2L^2}\left(\frac{\pr \rho}{\pr v}-\frac{\pr \lambda}{\pr v}\right)\left(\frac{du}{dS}\right)^2+\frac{\pr \lambda}{\pr u}\frac{dv}{dS} \frac{du}{dS}=0\label{geo7}\\
\frac{d^2 u}{dS^2}&+&\frac{L^2}{2l^2}\left(\frac{\pr \rho}{\pr u}-\frac{\pr \lambda}{\pr u}\right)\left(\frac{dv}{dS}\right)^2+\frac{1}{2}\left(\frac{\pr \lambda}{\pr u}+\frac{\pr \rho}{\pr u}\right)\left(\frac{du}{dS}\right)^2+\frac{\pr \lambda}{\pr v}\frac{dv}{dS} \frac{du}{dS}=0\label{geo8}
 \eea
\end{widetext}
Since the metric fields and their derivatives are all functions of $(u,v)$ coordinates, solving Eqs.(\ref{geo7},\ref{geo8}) 
requires integrating both ODEs while carefully interpolating the numerical fields $\lambda,\pr_{u,z}\lambda,\rho,\pr_{u,z}\rho$ at the current location point in each integration step. Eq.(\ref{lagrangian}) will be used as a constraint to validate the numerical integration.
%%%%% FIG  %%%%%%
\begin{figure*}[ht!]
\begin{tabular}{cc}
\includegraphics[trim={0cm 0 4cm 10cm},clip=true,scale=0.3]{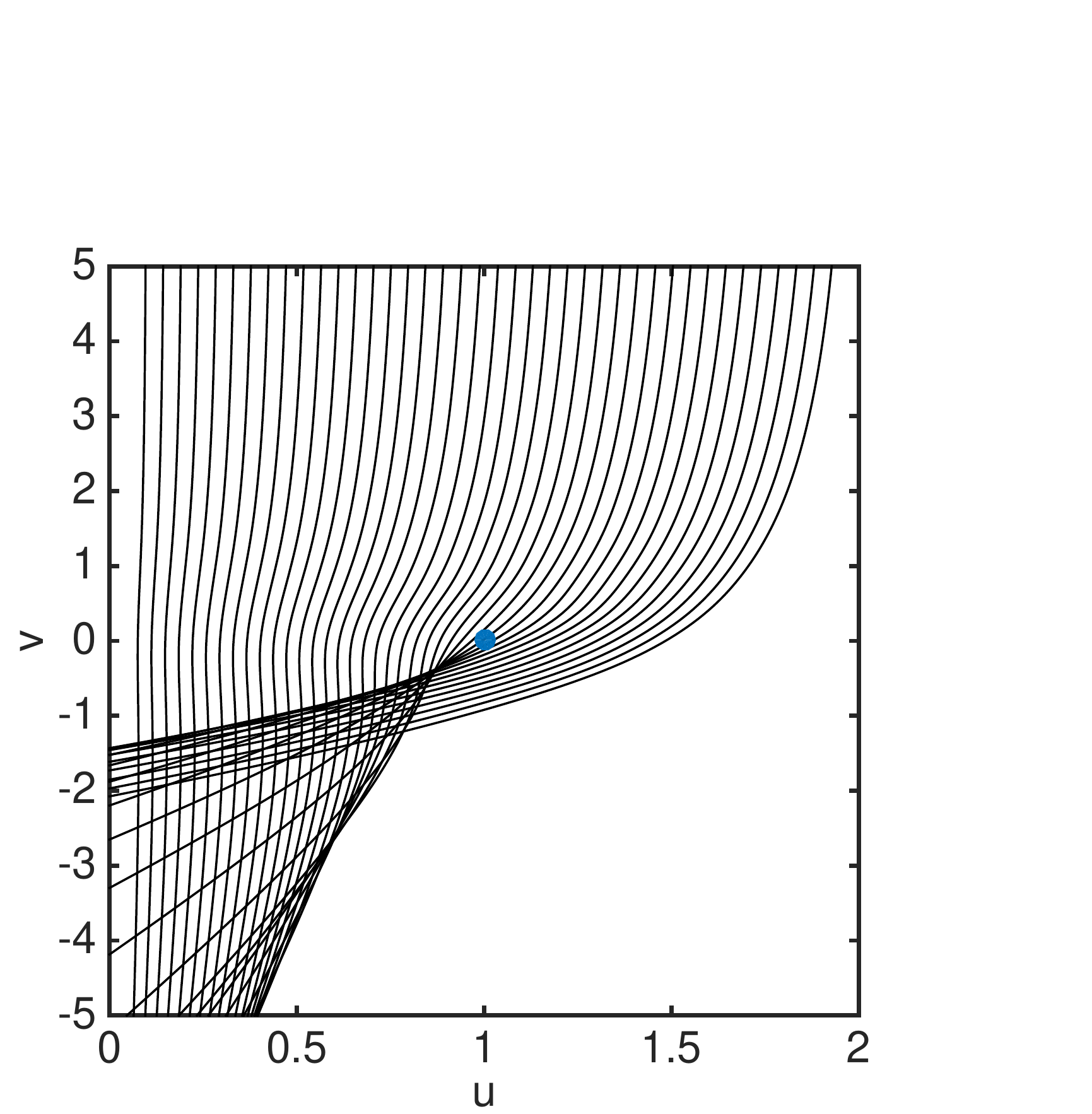}&
\includegraphics[trim={0cm 0 4cm 10cm},clip=true,scale=0.3]{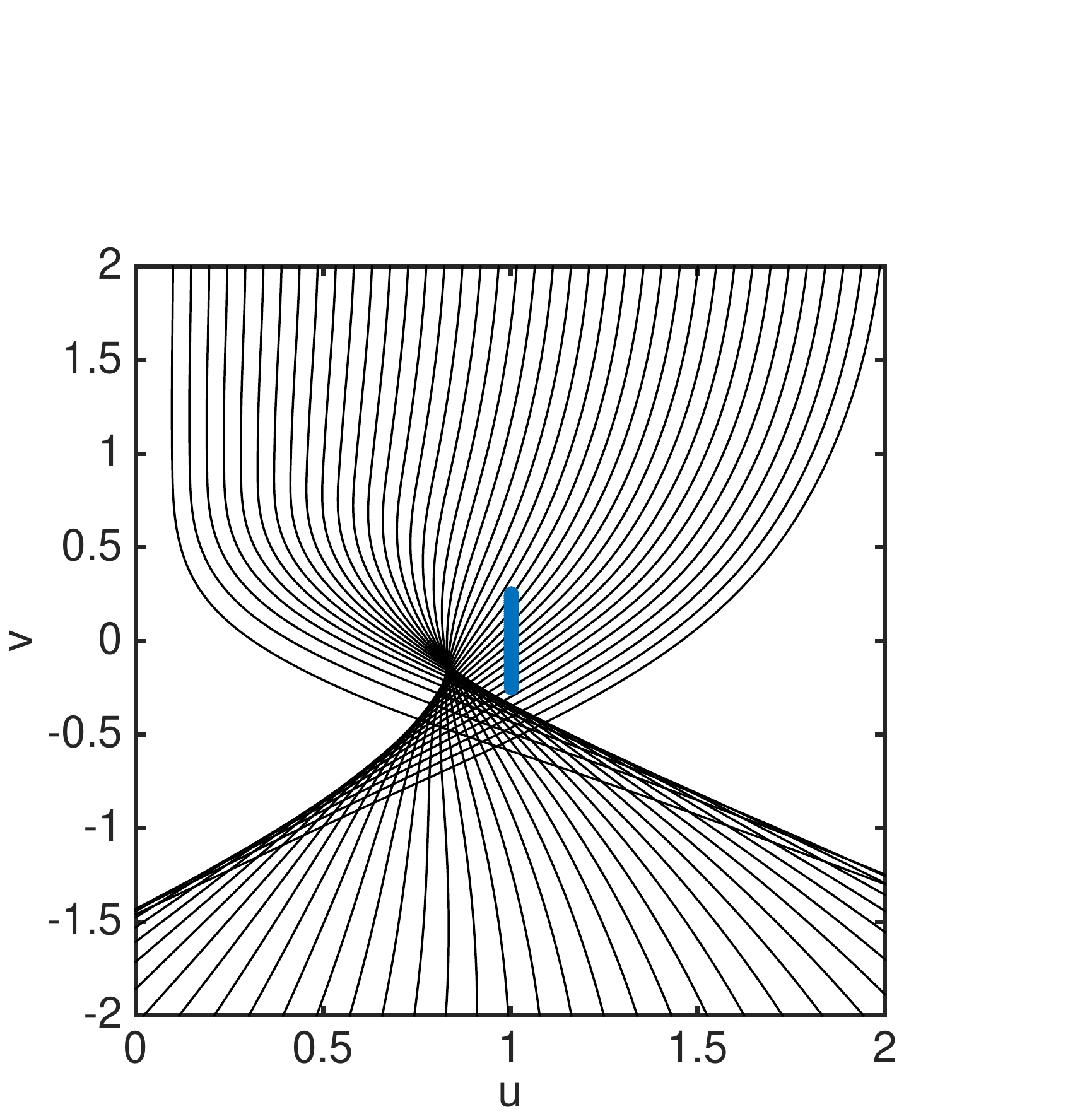} \\
\includegraphics[trim={0cm 0 4cm 10cm},clip=true,scale=0.3]{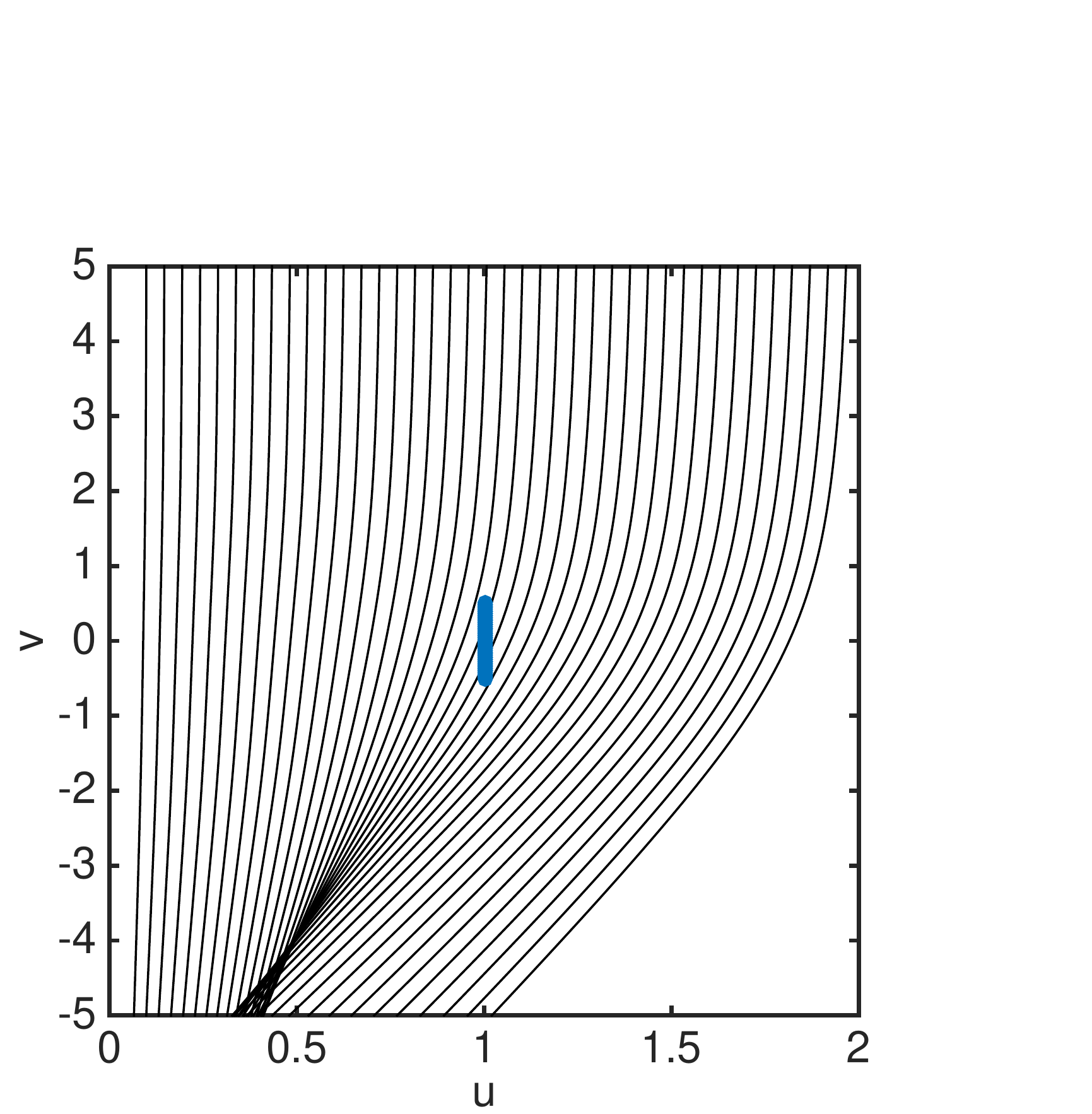}&
\includegraphics[trim={0cm 0 0cm 0cm},clip=true,scale=0.3]{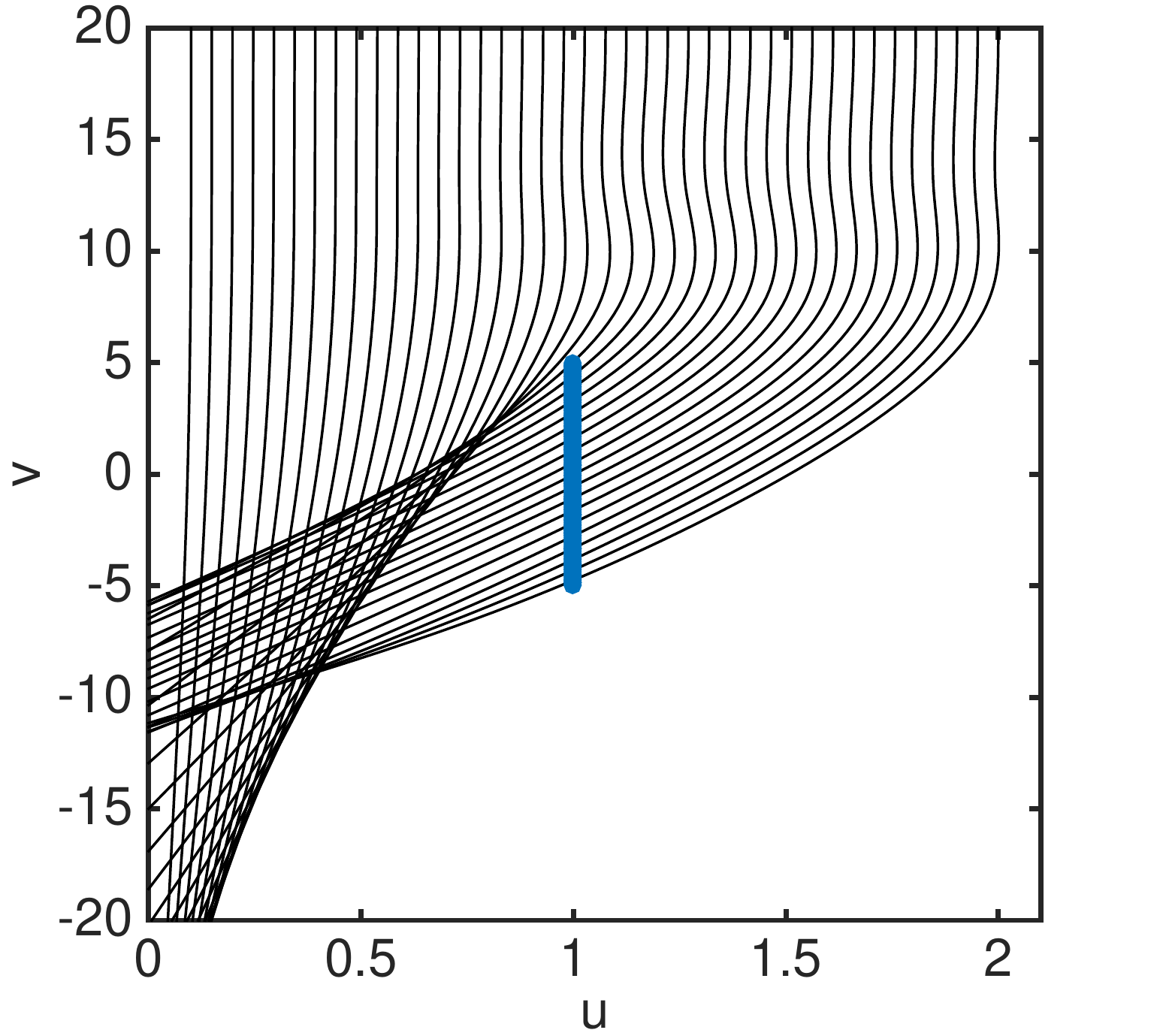}
\end{tabular}
\caption{Trajectories of parallel light rays coming
from infinity (top left panel: current loop ($\cI=10$),  top right panel: 
 solenoid with $L=0.5 l$ ($\cI=10$), lower left panel: solenoid with $L=l$ ($\cI=10$),  lower right panel: solenoid with $L=10l$ ($\cI=1$). Blue dots indicate the location of the current loop and the solenoids}\label{fig6}
\end{figure*}

Fig. \ref{fig6} presents the deflected trajectories of a bundle of light rays incoming parallel from spatial infinity $v\rightarrow+\infty$. These trajectories have been obtained from the numerical resolution of geodesic equations in strongly curved space-times around loop and solenoids with extremely large magneto-gravitational coupling $\cI=10$ (or $\cI=1$) so that the way light is deflected can be easily shown. The constraint Eq.(\ref{lagrangian}) is below $10^{-8}$
along the trajectories shown in Fig. \ref{fig6}. Geodesics passing far-away from the current loop of the solenoid follow hyperbola as if photons were attracted by a point mass. However, null geodesics with close encounters can exhibit more sophisticated shape
while passing through the magnetic device. In this case of strong coupling ($\cI=10$ or $\cI=1$), 
the deviation of the light beam is so strong that magnification appears at some locations while large regions of the $(u,v)$ plane have been cleared of any light. These results can be applied to the gravitational lensing of cosmic string loops in the strong field regime.
%%%%
\section{Application: Generation and Detection of Artificially Generated Gravitational Fields}
We now investigate how far such deflexion of light could be detected with the present technology of superconducting electromagnets and high precision light-wave interferometers. 
The basic idea is to make interfering two light beams among which one has travelled in the space-time curved by powered superconducting solenoids and the other not. The shorter distances travelled by the light beam inside the powered solenoids will generate a path difference between both light beams that will impact on their interference pattern as a result of a gravitationally generated phase shift. However, since the large electric currents that can be achieved with current superconducting cables, roughly of order $10^4A$, will generate extremely weak space-time curvature, it will be necessary to amplify the signal by forcing light to perform numerous round trips in the artificially generated gravitational field.

We can first write down the gravitational field equations in the weak field limit.
If we assume $a_{\rm rel}\ll a_{\rm nr}$ and $\rho,\lambda \ll 1$, we get 
that Eqs.(\ref{rho2}-\ref{lam2}) now reduce to
\bea
\nabla^2\rho&=&\frac{\cI}{u^2} \frac{L^2}{l^2}\left(\left(\pr_u a_{\rm nr}\right)^2+\frac{l^2}{L^2}\left(\pr_v  a_{\rm nr}\right)^2\right)\label{rho3}\\
\nabla^2\lambda &=&\frac{\cI}{u^2} \frac{L^2}{l^2}\left(\left(\pr_u a_{\rm nr}\right)^2-\frac{l^2}{L^2}\left(\pr_v  a_{\rm nr}\right)^2\right)\label{lam3}
\eea
where $\nabla^2=\pr_u^2 +\frac{1}{u}\pr_u+\frac{l^2}{L^2}\pr_v^2\cdot$ 
For reasons to be explained below, we will consider that the source of the magnetic fields is given by a set of stacked anti-Helmholtz coils (or multi-layered coils). The $i-$th anti-Helmholtz coil is constituted by two solenoids of radius $l_i$ ($l_i<l$), length $L$, spaced by a distance $D$ and carrying steady electric current of opposite directions.  It will be necessary to pile up these (anti-)Helmholtz coils to produce a detectable space-time curvature by means of electric currents of order $10^4 A$.
The corresponding expression for the
total magnetic potential $a_{\rm nr}$ of $n$ stacked anti-Helmholtz coils of radius $l_i$ can be obtained from Eq.(\ref{a_sol}) by
\be
a_{\rm nr}=\sum_{i=1}^n \left(a_{\rm nr}^{\rm sol}(r,z+D/2;l_i)-a_{\rm nr}^{\rm sol}(r,z-D/2;l_i)\right)
\ee
Corresponding boundary conditions can be obtained from those given in section II by linearly superposing the boundary conditions of single solenoids as allowed by weak gravitational field limit. The problem can be solved numerically with a standard spectral method (for instance based on Fourier decomposition, see section III).

\begin{figure}[ht!]
\includegraphics[scale=0.3]{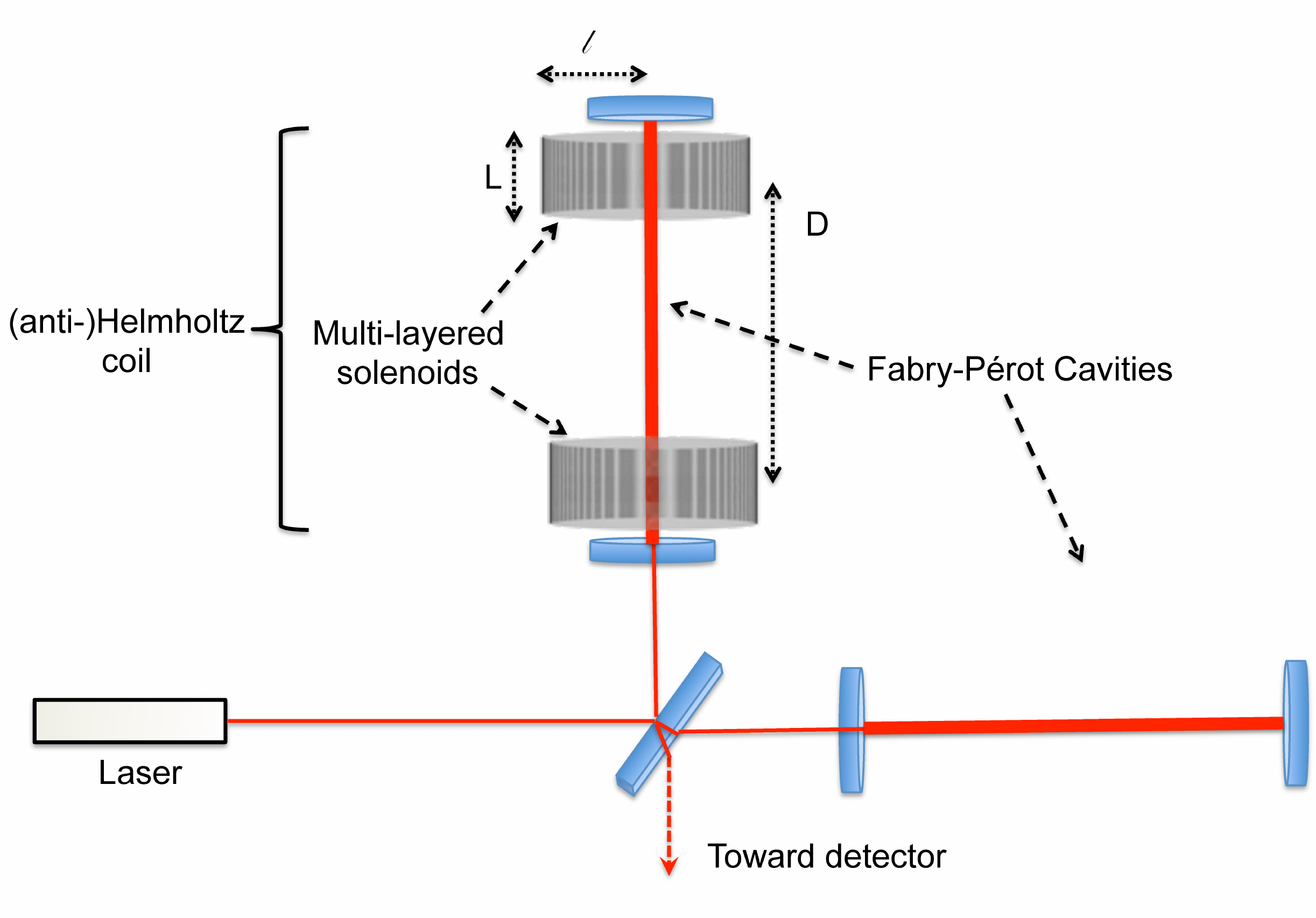} 
\caption{Schematic view of the proposed experimental set-up}\label{fig7}
\end{figure}
Our proposed experimental set-up, shown in Fig. \ref{fig7}, recalls those of ground-based interferometers used to detect gravitational waves: it consists of a Michelson interferometer whose arms are constituted by Fabry-P\'erot cavities. One of these arms  goes through a multi-layered anti-Helmholtz coil consisting of two stacks of superconducting solenoids. As long as the electric current is switched on in the device, this one curves space-time and deflects light. Since space-time is slightly shrunk inside the coil, light trapped inside the powered coil accumulates phase shift as the round trips succeed each other. 

Following \cite{stodolsky}, we can write down the phase shift due to the weak
gravitational field $h_{\mu\nu}$ on a Minkowski path $\gamma$ of light
\be
\Delta \Phi=\frac{1}{2}\int_\gamma h_{\mu\nu} K^{(0)\nu}dx^\mu
\ee
where $K^{(0)\nu}$ is the (unperturbed) constant 4-wave vector of the wave front.
We focus on light following the axis $u=0$ of the coil, with wave vector components
$K^{(0)t}=\frac{2\pi \nu}{c}=K^{(0)z}$ with $\nu$ the frequency of light. Therefore,
the phase shift along the axis of the coil for one trip is given by:
\be
\Delta \Phi=\frac{\pi}{\Lambda} \int_0^\mathcal{L} \left(\rho(0,z)-\lambda(0,z)\right) dz
\label{phase}
\ee
where $\Lambda=\frac{c}{\nu}$ is the wavelength of the light beam of frequency $\nu$ and $\mathcal{L}$
is the length of the interferometer arm. An anti-Helmholtz coil configuration is better than a solenoid for the production of gravitational phase shift. Indeed,  phase shift
is directly related to the difference $\rho-\lambda$ which obeys the following PDE (Eq.(\ref{rho3}) - Eq.(\ref{lam3})):
\be
\nabla^2\left(\rho-\lambda\right)=\frac{2\cI}{u^2} \left(\pr_v  a_{\rm nr}\right)^2\sim |B_r|^2,
\ee
where $B_r$ is the radial component of the magnetic field. This quantity, although vanishing on the axis of symmetry, increases more rapidly inside a Helmholtz coil than in the interior of a long solenoid, justifying the above-mentionned choice.
\\
As a matter of comparison, the phase shift induced by a gravitational wave passing by
a Michelson interferometer is given by \cite{barone}:
\be
\Delta \Phi= \frac{2\pi c}{\Lambda} |h| \tau_{\rm trip}
\ee
where $|h|$ is the amplitude of the incoming gravitational wave and $\tau_{\rm trip}=2\mathcal{L}/c$ is the single round trip travel time of the light beam inside the arm (to be  amplified 
through the multiple reflexions induced by the Fabry-Perot cavities). Currently achievable threshold for detection is for $|h|\approx 10^{-21}$, $\mathcal{L}\approx 10^3 m$
and $\Lambda\approx 10^{-6} m$ yielding $\Delta\Phi\approx 10^{-11}\rm rad$ per trip.

Let us now give an estimation of this phase shift for realistic experimental conditions of the setup presented above. 
We particularize the setup as following. We consider a set of 10 stacked anti-Helmholtz coils, each constituted by
two superconducting solenoids of same length $L=2.5 m$ carrying opposite steady electric current of $20kA$ (which is similar to CMS-class magnets \cite{cms}) spaced by a distance of $D=2.5 m$. The external solenoids have a radius of $l=5m$ and the 10 solenoid shells are chosen equally spaced between $r=1m$ and $r=5m$. The length of the interferometer arm has been chosen to $\mathcal{L}=50m$.

\begin{figure}[ht!]
\includegraphics[trim={0cm 4cm 0cm 4cm},clip,scale=0.3]{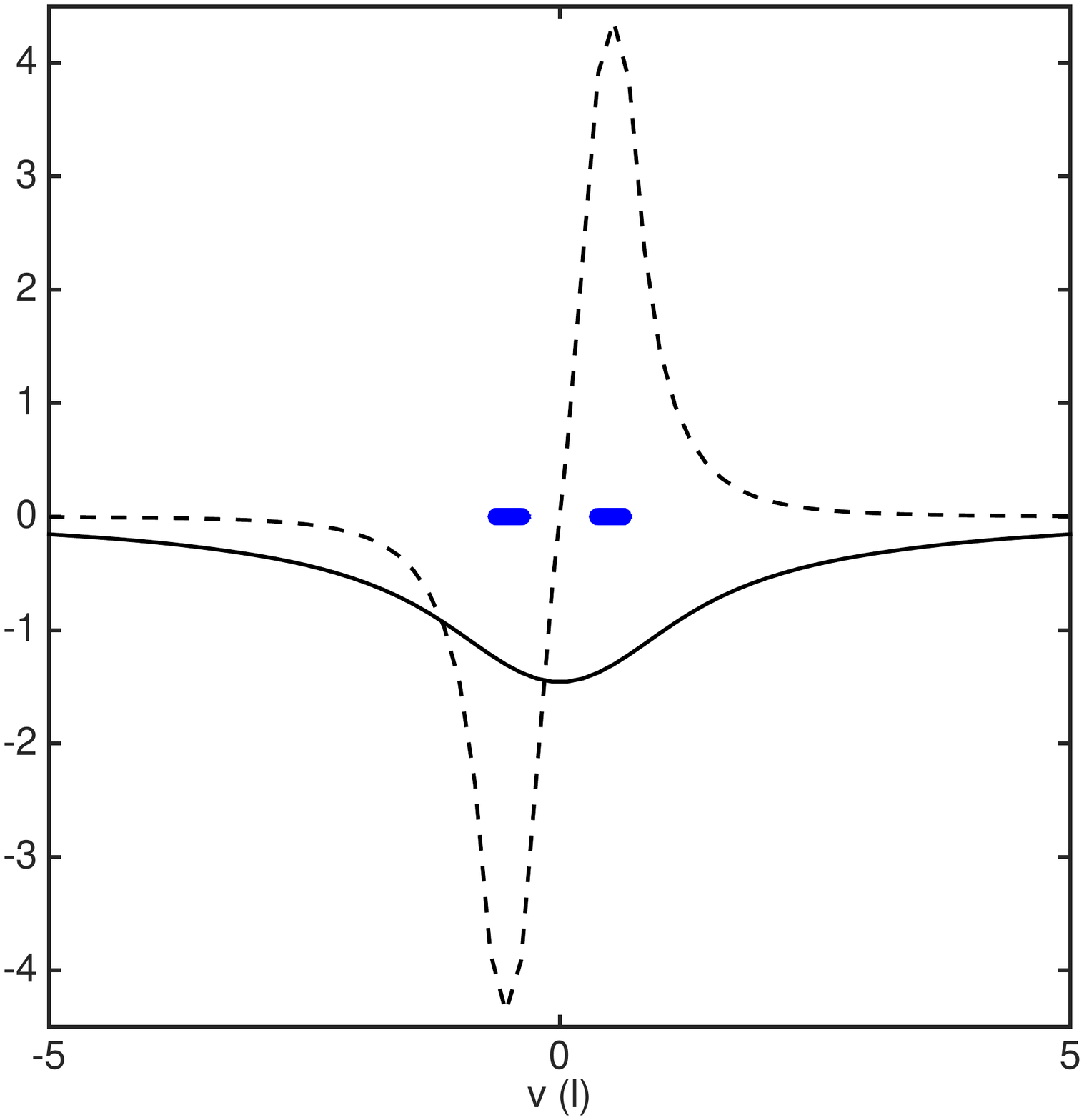} 
\caption{$(\rho(0,v)-\lambda(0,v))/\cI$ (straight line) and magnetic field $B_z(0,v)/(\mu_0 N I )$ (dashed line) along the axis of symmetry of the anti-Hemlholtz coil ($\mu_0 N I=4.4 T$). Dots indicate the position of the solenoids along the $z-$axis}\label{fig8}
\end{figure}

Figure \ref{fig8} shows the profiles of $B_z(0,v)$ and $\rho(0,v)-\lambda(0,v)$ along
the axis of symmetry $u=0$ of the solenoids. The anti-Helmholtz coils generates a curvature of spacetime that reaches its maximum at mid distance from each of the solenoids ($u=0$, $v=0$) where the magnetic field vanishes. The magnetic field is maximal at the center of the solenoids (at $z=\pm D/2$), and its magnitude reaches about $20 T$ for the parameters chosen above. The phase shift Eq.(\ref{phase}) can be integrated numerically for the results of Figure \ref{fig8} and we find $\Delta\phi\approx -1.56\times 10^{-25}$ (for $\Lambda=514 \rm nm$) per round trip inside the interferometer. If the experiment can be conducted long enough, for a time $T_{\rm exp}$, this phase shift will be accumulated. For two hundred days of duration, the accumulated phase
shift reaches $\Delta\phi\approx -1.08\times 10^{-11}$. This value is of the same order of magnitude to that of a gravitational wave signal\footnote{A gravitational wave coming from astrophysical sources however produces the same amplitude in
timescales of a millisecond.}, and could be detected by current technologies developed for ground-based gravitational waves observatories \cite{GW}. 

The example given above is purely indicative and aims to show that a detectable phase shift could be produced by present day technology. We can give a simple order of magnitude and lower bound of the accumulated phase shift produced by posing $\rho-\lambda\approx \cI$ on the axis of symmetry so that the phase shift Eq.(\ref{phase}) gives 
\be
\Delta\Phi\approx \pi\frac{\mathcal{L}}{\Lambda}\cI \times \frac{T_{\rm exp}}{t_{\rm bounce}}\approx \pi\frac{\cI}{\Lambda}c T_{\rm exp}
\ee
where $T_{\rm exp}$ is the duration of the experiment and $t_{\rm bounce}$ is the time taken
by light to produce a bounce inside the interferometer so that $T_{\rm exp}/t_{\rm bounce}=N_{\rm bounce}$ is the number of bounces inside the Fabry-Perot cavity.

To conclude this section, we emphasize that the generation and detection of artificial gravitational fields by strong magnetic fields is within experimental reach but requires
large multi-layered superconducting magnets powered during dozens of days as well as hundred meters long Michelson interferometers with Fabry-Perot cavities that could achieve the same sensitivity than ground-based gravitational wave observatories but in presence of intense magnetic fields. 
%% add response to referee comment (c) here
As in gravitational wave observatories, a long optical path is crucial for the detection of the very weak gravitational perturbations. However, the wavefront of an astrophysical gravitational wave is much longer than the perturbations of space-time generated by  electro-magnets a decameter large. Therefore, a kilometer-wide interferometer can fit into  the gravitational wavefront coming from astrophysical sources and is necessary to capture the wave during the time it passes through Earth. Using a kilometer-large interferometer for the detection of the space-time curvature induced by superconducting coils would require kilometer-large magnets. This is not necessary since, at the opposite of the detection of gravitational waves, space-time deformation by electro-magnets is maintained as long as the magnetic field is present. The amplitude of this space-time deformation is extremely tiny, of order of $\cI$, which requires to trap the light long enough inside the space-time deformation to accumulate enough phase shift for detection. 
Although experimentally challenging, such a detection would open the path to a new class of laboratory tests of general relativity and the equivalence principle. 

\section{Conclusions}
The generation of artificial gravitational fields with electric currents could be in principle detected through the induced change in space-time geometry that results in a purely classical deflexion of light by magnetic fields. This effect does not invoke any new physics, as it is a consequence of the equivalence principle. Although very weak, we have shown that this effect could be detectable by a twofold experimental setup. On one hand, it includes stacked large superconducting Helmholtz coils for the generation of the artificial gravitational field. On the other hand, the detection would be achieved by highly sensitive Michelson interferometers whose arms contain Fabry-Perot cavities to store light into the generated gravitational field. In an appropriate experimental set-up, the amplitude of
the phase shift accumulated during the bouncing of light in the curved space-time generated
by the magnetic field would reach in a few months the level of an astrophysical source of gravitational wave passing through ground-based GW observatories.

We claim that such detection would open new eras in experimental gravity and laboratory tests of general relativity and the equivalence principle. These tests, although concerning the weak field regime, will have the particularity of focusing exclusively on the coupling between gravitation and electromagnetism. Future theoretical works should focus on extending the present study to alternative theories of gravity to explore how far they would depart from general relativity. 

Such a detection of the space-time curvature generated by a magnetic field in laboratory would constitute a major step in physics: the ability to produce, detect, and ultimately control artificial gravitational fields. And would this technology be developed, it could lead to amazing applications like the controlled emission of gravitational waves with large alternative electric currents. Gravity would then cease to be the last of the four fundamental forces not under control by human beings.

\bigskip
\noindent \textit{Acknowledgments:} The author is very grateful to M. Rinaldi and A. Hees for the fruitful discussions which significantly helped to extend the preliminary work and ended up with the results presented as an application. All computations were performed at the ``plate-forme technologique en calcul intensif'' (PTCI) of the U. of Namur, Belgium, with financial support of the F.R.S.-FNRS (convention No. 2.4617.07. and 2.5020.11).

\end{document}